\documentclass[aps,prb,twocolumn,groupedaddress]{revtex4-1}\usepackage{hyperref}
\pdfoutput=1

\usepackage{amsmath}
\usepackage{amssymb}
\usepackage{bm}
\usepackage{graphicx}

\usepackage{hyperref}
\hypersetup{%
  breaklinks = {true},
  citecolor = {blue},
  colorlinks = {true},
  linkcolor = {red},
  pdfauthor = {\textcopyright\ Matteo Bazzanella},
  pdfcreator = {\LaTeX\ and \flqq hyperref\frqq},
  pdffitwindow = {true},
  pdfmenubar = {true},
  pdfpagelayout = {SinglePage},
  pdfstartview = {Fit},
  pdftoolbar = {true},
  plainpages = {false},
}

\newcommand{\la}{\langle}
\newcommand{\ra}{\rangle}
\newcommand{\ua}{\uparrow}
\newcommand{\da}{\downarrow}
\newcommand{\dd}{\dagger}
\newcommand{\be}{\begin{eqnarray}}
\newcommand{\ee}{\end{eqnarray}}
\newcommand{\boldgreek}[1]{\ensuremath{\mbox{\boldmath$#1$}}}
\usepackage{color}
\usepackage{color}

\begin{document}


\title{Ferromagnetism in the one-dimensional Kondo lattice:\\mean-field approach via Majorana fermion canonical transformation}

\author{Matteo Bazzanella and Johan Nilsson}
\address{Department of Physics, University of Gothenburg, 412 96 Gothenburg,  Sweden}

\date{November, 2013}

\begin{abstract}

Using a canonical transformation it is possible to faithfully represent the Kondo lattice model in terms of Majorana fermions. Studying
this representation we discovered an exact mapping between the Kondo lattice Hamiltonian and a
Hamiltonian describing three spinless fermions interacting on a lattice.
This alternative form of the Hamiltonian is suitable for an immediate identification of the competing effects that operate in the Kondo lattice model.
In particular a term describing the double exchange mechanism appears explicitly in the Hamiltonian.
We investigate the effectiveness of this three fermion representation by performing a zero temperature mean-field study of the
phase diagram at different couplings and fillings for the one-dimensional case, focusing on the appearance of ferromagnetism.
The solutions show interesting features that agree in many respects with the
known numerical and analytical results.
In particular, in the ferromagnetic region connected to the solution at zero electron density, we have a quantitative agreement
on the value of the ``commensurability parameter'' discovered in recent DMRG (in one dimension) and DMFT (in infinite dimensions)
simulations; furthermore we provide a theoretical justification for it, identifying a symmetry of the Hamiltonian.
This ferromagnetic phase is stabilized by the emergence of a spin-selective Kondo insulator that is
described quite conveniently by the three spinless fermions. We discover however that such a phase
cannot be the correct description for all the ferromagnetic phases of the one-dimensional Kondo lattice model. We found in fact
a different ferromagnetic phase at high filling and low couplings. This phase resembles the RKKY ferromagnetic phase existing at vanishing filling,
but it incorporates much more of the Kondo effect, making it energetically more favorable than the typical spiral (spin ordered) mean field ground states.
We believe that this second phase represents a prototype for the strange ferromagnetic tongue identified by numerical simulations
inside the paramagnetic dome.
At the end of the work we also provide a discussion of possible orders different from the ferromagnetic one. In particular at half-filling, where
we obtain as ground state at high coupling the correct Kondo insulating state.
\end{abstract}

\maketitle

\section{Introduction}
\label{sec:intro}

The Kondo lattice is one of the most studied models in condensed matter physics\cite{Tsunetsugu:1997vn,Gulacsi-:2004nx,Gulacsi:2006oq,Hewsonbook}. 
The features of this model are approximately explained by the competition between three main phenomena: the
coherent formation of Kondo-singlets\cite{JPSJ.74.4,Burdin:2000bh,Meyer:2000dq},
the RKKY spin-spin effective interaction\cite{Fazekas,Yosida:1957ud,Ruderman:1954if} and the double
exchange mechanism\cite{Anderson:1955fk,Zener:1951uq,Gennes:1960kx}.
The balance between these effects gives rise to highly non-trivial physics that is not easily described keeping
the impurity spin and conduction electron degrees of freedom strictly separated.
As a consequence the phase diagram of the Kondo Lattice model (KLM) is quite fascinating: by changing only two parameters
(the Kondo coupling and the density of the conduction electrons) many
different classes of ground states can be explored\cite{Tsunetsugu:1997vn,Gulacsi-:2004nx,Gulacsi:2006oq,Hewsonbook}. 
The KLM can therefore be used as the model Hamiltonian for many interesting systems, such as
heavy fermion compounds\cite{Coleman:1983ys,Coleman:qf,Ueda:1997zt} 
and Kondo insulators\cite{Le-Hur:1998ly,Tsunetsugu:1997vn,Eder:1998fu,Eder:1997kl,Tian:1994pi,Tsvelik:1994hs}.
It is also believed that many of the currently most studied systems (both by numerical and experimental means), for example actinide
compounds and perovskities\cite{Tsunetsugu:1997vn,Gulacsi-:2004nx,Gulacsi:2006oq,Hewsonbook}, can be effectively described by the KLM.
The existence in the KLM of superconductivity generated by spin-fluctuations has also been a source of debate
and has recently been addressed by new numerical studies\cite{Bodensiek:2013fk,Fabrizio:2013kondoSC} that
suggested the presence of this mechanism in the Kondo-Heisenberg model.

Unfortunately few exact solutions of this model are available, and typically only limited regions of the phase diagram can be
successfully described by analytical tools.
The problems are of course created by the interaction between the conduction electrons and
the impurity spins that causes the entanglement\cite{Sorensen:2007nu,*Eriksson:2011sw} of these
two quantum degrees of freedom.
This poses formal problems and raises profound questions:
the former are due to the different operator algebras describing the conduction electrons and the impurity spins
that cannot be treated on the same footing; the latter
are related to the role of the impurity spins and their debated
contribution to the Fermi volume\cite{Ueda:1994kl,*Coleman:2001hc,*Senthil:2003ij,*Lohneysen:2007bs,*Tsvelik:2001fv,*Yamanaka:1997dz}.

In some particular limits the one-dimensional Hamiltonian can be studied analytically.
For example some exact results exist at half-\cite{Le-Hur:1998ly,Tsunetsugu:1997vn,Eder:1998fu,Eder:1997kl,Tian:1994pi,Tsvelik:1994hs}
and infinitesimal-filling\cite{Sigrist:1991fu}
for every value of the Kondo coupling, or at infinite\cite{Lacroix:1985lh} and high\cite{Sigrist:1992qa} Kondo coupling for every value of the filling.
Moreover the low energy properties of the system can be solved exactly for weak Kondo couplings,
making use of bosonization techniques\cite{Le-Hur:1998ly,Honner:1997qo,Le-Hur:2000tw,JPSJ.63.4322}.
However, to explore larger regions of the phase diagram, only numerical methods\cite{Tsunetsugu:1997vn,Yu:1993jl,Troyer:1993gb,Shibata:1999mb,McCulloch:2002ss}
can be applied. 
A standard approach, common to many numerical and analytical studies,
is to start from the analysis of the Anderson lattice model and strongly enforce
unit-occupancy of the $f$-impurity states.
This generates a constraint on the local number of the $f$-electrons that, if fulfilled exactly, implies the freezing of their
charge degree of freedom, transforming effectively the $f$-electrons into impurity spins\cite{Coleman:qf,Meyer:2000dq}.
Such an approach to the problem relies on the fact that the Kondo lattice model can be seen as
the effective low energy description of the symmetric Anderson lattice model\cite{Lacroix:1979aa,Fazekas:1991kc,Hagymasi:2012qc}, if the hybridization
and the confinement potential are appropriately sent to infinity, as prescribed by the Schrieffer-Wolff transformation\cite{Schrieffer:1966mw,*Sinjukow:2002bd}.
The positive aspect of this approach is that all the excitations of the system are fermionic; the constraint is however
difficult to implement, so generally it can be fulfilled only on the average. 
Although this inexact realization of the constraint can be meaningful in the
study of the properties of real materials, it can also encode some unwanted features. For example, if the subject of the study is
spin mediated superconductivity, it is not clear wether or not the analysis will be spoiled by the 
valence fluctuations of the $f$-impurity electrons.
Moreover, as the Kondo lattice is an effective description of the Anderson lattice,
it would be reasonable to be able to solve it without any reference to
the high-energy physics that is integrated out by the Schrieffer-Wolff procedure.
Only in this way can one expect to learn something about the low energy physics of the system
and the driving mechanisms and interactions that are relevant at this low energy scale.

From this point of view we believe that it could be fruitful to represent the Kondo lattice Hamiltonian
in terms of properly defined fermionic degrees of freedom: a formulation of the problem that {\it requires no implementation
of the constraint, but is given anyway in terms of fermions only}.
In this paper we elaborate and discuss such a representation, identifying the map between the old degrees of freedom
(impurity spins and conduction electrons) to the new ones (three spinless fermions $c$, $g$ and $f$).
For sake of clarity we will refer to this transformation as the $cgf$-map in the rest of this manuscript. It is important to
remark on the fact that our approach holds only for the KLM with spin-1/2 impurities.

The mapping could simply be defined and proved by brute force calculation, but
we will also show the path that we followed to discover it, believing that a more pedagogical approach will make
our method more interesting. The derivation
goes through the representation of the electron and spin operators in terms of their basic Majorana fermion constituents.
Writing the Kondo lattice Hamiltonian in terms of Majorana fermions\cite{JohanKondoORIGINAL}, we treat the electrons and the spins on equal footing,
avoiding the most
problematic feature of the Hamiltonian.
Our $cgf$-map is a practical example of the usefulness of this Majorana fermion based approach.



Although the map holds for any number of dimensions and any lattice structure, we will study only the one-dimensional case at zero temperature, in
order to have a comparison with the existing literature. We will perform a mean-field, zero-temperature analysis of the KLM,
represented on the set of the three spinless fermions.
It will become clear how our treatment is able to naturally include, {\it already at the mean-field level},
all the three main mechanisms of RKKY interaction, local Kondo-singlets formation and double exchange.
In particular our approach very conveniently describes the latter two effects, permitting a mean-field analysis of the
ferromagnetic regions of the phase diagram, on which we focus our attention.

Considering the length of this manuscript and the large amount of physics that we are going to discuss, we introduce here briefly the results of the mean-field
analysis that we hope will convince the reader of the relevance of our work.

Our mean-field results are in good agreement with
the picture recently provided by DMRG\cite{Peters:2012fc} and DMFT\cite{Peters:2012rq} studies of the ferromagnetic
metallic phase for low to intermediate values of the Kondo coupling.
In fact we identify a phase (FM-I) where we obtain a {\it quantitative agreement} on the values of the average on-site total magnetization and
of the ``commensurability parameter''. For the latter we provide also a theoretical justification, identifying the symmetry operation that enforces it.
In the same phase we recognize the fundamental role of the double exchange mechanism, which
allows for the generation of the ``spin-selective Kondo insulator'' (SSKI)
and the separation of the electrons in majority- and minority-electrons. While the majority-electrons behave as normal non-interacting electrons, the minority-electrons
appear only as constituents of delocalized Kondo-singlets, as made explicit by the disappearance of their Fermi surface.
The generation of the SSKI stabilizes the FM-I phase, but this does not mean that ferromagnetic order needs the realization of the SSKI.
In fact for low couplings we discovered that the ferromagnetic phase can extend up to half filling, but for a
coupling dependent critical density $n^{F}_{crit}$ the mechanism stabilizing
the ferromagnetic order of the impurity spins changes completely, marking the emergence of a new phase FM-II.
The transition from the FM-I and FM-II phases
happens approximately at the known\cite{McCulloch:2002ss} phase-separation line that divides the ferromagnetic metallic phase and the paramagnetic phase.
We believe that the FM-II phase is an improvement with respect to the RKKY-ferromagnetic state, able to incorporate the Kondo effect in a more efficient way. This property
permits the FM-II phase to survive up to half filling, if the coupling is not too strong. This phase is energetically competitive, if compared to the usual
spiral ordered mean-field trial\cite{Lacroix:1979aa,Fazekas:1991kc} ground-states.
In the light of these features the FM-II state is a natural candidate
for the description of the ferromagnetic ``tongue'' phase identified\cite{McCulloch:2002ss} inside the paramagnetic dome by DMRG simulations.


For intermediate couplings an instability appears and the ferromagnetism of FM-II ceases exist. In these situations,
another coupling dependent critical density $n^{pol}_{crit}$ appears for the FM-I phase. Above this density we found
no translational invariant mean field solutions, except the half-filled one. In proximity of $n^{pol}_{crit}$ we note that the compressibility
of the FM-I phase tends to infinity, so we interpret this feature as an indication of the fact that the two componets
forming the FM-I state (the majority-electrons and Kondo-singlets liquids)
start to separate. This process creates in the phase diagram
a region of phase coexistence between the FM-I and
the half-filled Kondo Insulating (KI) phase, where the latter is characterized by the absence of free electron modes.
The location in the phase diagram and the physical picture of this region, in correspondence of the disappearance of the FM-II phase,
are in good agreement with the description of the
polaronic liquid provided by bosonization\cite{Honner:1997qo,Gulacsi-:2004nx,Gulacsi:2006oq}.
In terms of the qualitative two liquid picture mentioned previously, the polarons can be identified as the ``islands'' of ferromagnetic FM-I phase
immersed in the liquid of Kondo singlets of the half-filled KI phase.

Increasing even more the coupling the solutions become less and less meaningful. We believe that this is a symptom of the fact that the
assumptions behind our mean-field approximation are not justified anymore.

Unfortunately our mean field decomposition scheme is not very convenient for the study of different kinds of magnetic orders,
because the form of the Kondo Hamiltonian in terms of fermions $c$, $g$ and $f$ is naturally suited for
solutions that are translationally invariant. Consequently we decided to not perform an analysis of the
RKKY-liquid phase\cite{McCulloch:2002ss,Tsunetsugu:1997vn}, believing
that a different kind of approach would be more convenient\cite{ChristmaMajorana}.
Instead we briefly discuss the half-filled case, though we refer to the literature
for a more appropriate approach\cite{JohanKondoORIGINAL,ChristmaMajorana}, still based on the Majorana representation of the Kondo lattice Hamiltonian.
At half-filling we found a translational invariant KI solution, composed by a coherent superposition of local Kondo singlets;
this solution becomes an energetically favorable ground state, at high couplings,
if compared with the usual Ne\'{e}l ordered one. These features are
consistent with the known results\cite{Le-Hur:1998ly,Tsunetsugu:1997vn,Eder:1998fu,Eder:1997kl,Tian:1994pi}.
This KI ground state tends for $J\rightarrow +\infty$ to the correct ground state, where on each site a Kondo singlet is formed; moreover
it disappears if a $J$-dependent critical chemical potential $\mu^{pol}_{crit}(J)$ is reached. This chemical potential is the energy necessary to
remove an electron from the system, therefore it must correspond to the quasiparticle gap\cite{Yu:1993jl,Tsunetsugu:1997vn}.
The evolution of $\mu^{pol}_{crit}(J)$ with the coupling agrees very well with the results obtained via perturbative approaches\cite{Tsunetsugu:1997vn},
at high coupling, but does not share the same behavior away from this limit.



Showing the wide range of physics that can be described by following our approach to the problem, made possible by the
representation of the Hamiltonian in terms of Majoranas, we hope to convince the reader
that it could be interesting and profitable to
tackle some of the open problems in condensed matter physics using the same line of thought.
Moreover we hope to provide the community with a new convenient starting point for the study of the KLM and in particular the study
of its ferromagnetic properties in one and many dimensions.

The paper is organized as follows: in Sec.~\ref{sec:majoranamap} we represent faithfully the Kondo lattice
Hamiltonian in terms of Majorana fermions\cite{JohanKondoORIGINAL} and review the known\cite{MatsStellanHUBBARD,KumarORIGINAL} typical
properties common to these kind of mappings. Starting from this different formulation of the Hamiltonian we build the set of three
spinless-fermions in Sec.~\ref{sec:cgfmap}. 
After a discussion of the properties and the meaning of the different spinless fermions,
we will proceed in Sec.~\ref{sec:HartreeFock} to the mean-field analysis of the phase-diagram,
focusing on the one-dimensional case at zero temperature.
At the end we provide also a final outlook.

\section{Majorana map}
\label{sec:majoranamap}

In a recent work\cite{JohanKondoORIGINAL} it has been proved how the Kondo lattice Hamiltonian
\be
\label{KondoHamiltonian}
H_K&=&-t\sum_\sigma \sum_{\la i,j\ra} c^\dd_{c,\sigma}(\mathbf r_i)c_{c,\sigma}(\mathbf r_j)+\text{h.c.}\\
&&-\mu^*\sum_i n_c(\mathbf r_i)+J\sum_i \mathbf S_c(\mathbf r_i)\, \mathbf S_f (r_i),\nonumber
\ee
can be rewritten in terms of six Majorana fermion degrees of freedom (Majoranas). In the previous equation $J$ is the Kondo coupling,
$S_c$ and $S_f$ are respectively the electron- and impurity-spin operators,
$n_c$ is the conduction electron operator, $\mu^*$ is the Lagrange multiplier of the density constraint (i.e., the chemical potential) and $\la i,j \ra$
indicates the usual sum over nearest neighbors.

To understand the Majorana formulation, it is enlightening to start from the symmetric 1-site (local) Anderson impurity model
and analyze its Fock space. The Hamiltonian is given by:

\be\label{eq:andersonhamiltonian}
H_{1A}=-V\sum_{\sigma=\ua,\da}(c^\dd_{c,\sigma}c_{f,\sigma}+c^\dd_{f,\sigma}c_{c,\sigma})+U(n_f-1)^2,
\ee
where the subscripts $c,f$ denote two different fermion species (two different orbital indices) and $U,V$ are real parameters.
It is well known\cite{Schrieffer:1966mw,Meyer:2000dq,Coleman:qf,Hagymasi:2012qc} that in the
limit $U,V\rightarrow+\infty$, with $J=4V^2/U$, the Hamiltonian in Eq. (\ref{eq:andersonhamiltonian}) generates exactly
the spin-spin interaction term in (\ref{KondoHamiltonian}); hence the total Hilbert space
of the local (one-site) Kondo lattice model can be interpreted as the low energy 8-dimensional subspace of the original
16-dimensional local Anderson Fock space. The energy separation of the two subspaces
is due to the the interaction term $U(n_f-1)^2$ that brings no corrections to the energy for $n_f=1$, while it gives a contribution proportional to $U$ in case
$n_f=0$ or $2$. For $U\rightarrow +\infty$ the states with $n_f=0,2$ become inaccessible, so the states that span the low energy Hilbert space are those with one $f$
electron per site. This also means that the $f$-electron charge oscillations in the system are infinitely suppressed.
As a consequence the Kondo lattice model can be thought of as an Anderson impurity lattice model where the $f$ electron
density obeys the {\it exact} local constraint $n_f=1$.

The Majorana fermion description of the Kondo lattice model\cite{JohanKondoORIGINAL} stems from these considerations,
but implements them in a completely different way. Starting from the 1-site Anderson Hamiltonian
it is possible to set up a non-linear canonical transformation\cite{StellanKONDO} that separates the local Hilbert space in two sectors of low and high energy.
The connection between the two spaces
is given by a fermion operator $c_4^\dd$,
whose density is the only operator proportional to $U$ in the Hamiltonian.
The local Kondo Hilbert space is then
given by the states that contain no $c_4$-fermion.
It is thus possible to show\cite{JohanKondoORIGINAL} that the Hamiltonian (\ref{KondoHamiltonian}) can be rewritten using Majorana fermion degrees of freedom (Majoranas).
We define the Majoranas\cite{Coleman:1993wb,Coleman:1995oa,Shastry:1997xz}
in terms of our original operators in (\ref{KondoHamiltonian}) via
\be
c^\dd_{c,\ua}(\mathbf r_i)&=&\frac{\gamma_1(\mathbf r_i)+i\gamma_2(\mathbf r_i)}{\sqrt{2}},\label{eq:cupmajorana}\\
c^\dd_{c,\da}(\mathbf r_i)&=&\frac{-\gamma_3(\mathbf r_i)+i\gamma_4(\mathbf r_i)}{\sqrt{2}},\label{eq:cdownmajorana}
\ee
\be
&S_f^x (\mathbf r_i)=-i\mu_2(\mathbf r_i)\mu_3(\mathbf r_i),\quad S_f^y(\mathbf r_i)=-i\mu_3(\mathbf r_i)\mu_1(\mathbf r_i),& \nonumber\\
\\
&S_f^z(\mathbf r_i)=-i\mu_1(\mathbf r_i)\mu_2(\mathbf r_i),\label{eq:kumarspins}\label{eq:kondospins}& \nonumber
\ee
with the convention for the Clifford algebra of the Majorana operators: $\left\{\alpha_i,\beta_j\right\}=\delta_{i,j}\delta_{\alpha,\beta}$,
that means $\alpha_i^2=1/2$, where $i,j=1,2,3$ and $\alpha$, $\beta$ can be both $\gamma$ and $\mu$.

The faithful representation of the Kondo lattice Hamiltonian is obtained replacing the Majorana $\gamma_4(\mathbf r_i)$ with
the $\gamma$-independent Majorana $\gamma_0(\mathbf r_i)=2i \mu_1(\mathbf r_i)\mu_2(\mathbf r_i)\mu_3(\mathbf r_i)$, so that
$$
c^\dd_{c,\da}(\mathbf r_i)=\left(-\gamma_3(\mathbf r_i)+i\left[2i \mu_1(\mathbf r_i)\mu_2(\mathbf r_i)\mu_3(\mathbf r_i)\right]\right)/\sqrt{2}.
$$
In these terms the Kondo Hamiltonian (\ref{KondoHamiltonian}) is re-casted as
\begin{widetext}
\be\label{MajoranaHamiltonian}
H_{M}&=&-it\sum_{n,\mathbf\delta}\Big\{\gamma_2(\mathbf r_n)\gamma_1(\mathbf r_n+\boldgreek{\delta})-\gamma_1(\mathbf r_n)\gamma_2(\mathbf r_n+\boldgreek{\delta})
+\gamma_3(\mathbf r_n)\gamma_0(\mathbf r_n+\boldgreek{\delta})-\gamma_0(\mathbf r_n)\gamma_3(\mathbf r_n+\boldgreek{\delta})\Big\}+\\
&& \qquad +\frac{J}{4}\sum_n \Big\{i\gamma_1(\mathbf r_n)\mu_1(\mathbf r_n)+i\gamma_2(\mathbf r_n)\mu_2(\mathbf r_n)+i\gamma_3(\mathbf r_n)\mu_3(\mathbf r_n)+2\Bigl(\gamma_2(\mathbf r_n)\mu_2(\mathbf r_n)\gamma_3(\mathbf r_n)\mu_3(\mathbf r_n)+\nonumber\\
&&\qquad \qquad\qquad\qquad\qquad \qquad \qquad \qquad+\gamma_1(\mathbf r_n)\mu_1(\mathbf r_n)\gamma_3(\mathbf r_n)\mu_3(\mathbf r_n)+\gamma_1(\mathbf r_n)\mu_1(\mathbf r_n)\gamma_2(\mathbf r_n)\mu_2(\mathbf r_n)\Bigl)\Big\}+\nonumber\\
&& \qquad -\mu^* \sum_n \Bigl\{ -i\gamma_1(\mathbf r_n)\gamma_2(\mathbf r_n) -i\gamma_0(\mathbf r_n)\gamma_3(\mathbf r_n)+1\Bigl\},\nonumber
\ee
\end{widetext}
where $\mathbf r_n$ is summed over every lattice site and $\boldgreek{\delta}$ are the Bravais lattice vectors; so for example
in the one dimensional case $\sum_{n,\mathbf\delta}\gamma_2(\mathbf r_n)\gamma_1(\mathbf r_n+\boldgreek{\delta})=\sum_{n}\gamma_2(\mathbf r_n)\gamma_1(\mathbf r_{n+1})$.

This reformulation of the Kondo lattice Hamiltonian, although derived from the Anderson picture (\ref{eq:andersonhamiltonian}), has the major
advantage of being constraint-free and of treating the degrees of freedom of both
the electron and the localized spin on equal footing.
To the best of our knowledge there exists no other formulation of the Kondo lattice model that realizes these two features simultaneously.

The reader should keep in mind that the Majorana $\gamma_0$ is composed by the three Majoranas $\mu_1,\mu_2,\mu_3$.
In a recent work\cite{ChristmaMajorana} we analyzed
this dual nature of the Majorana $\gamma_0$, suggesting that this Majorana fermion should be the correct degree of freedom to describe the system
for small values of the coupling $J$. However in the present work we take a different approach, and we will instead consider $\gamma_0$
as a short-hand notation to indicate the three-body object $2i\mu_1\mu_2\mu_3$.

The proof of Eq.~(\ref{MajoranaHamiltonian}) has been already outlined\cite{JohanKondoORIGINAL}, but
thanks to the identification of the $cgf$-map we will be able to provide a more straightforward derivation of it, in this manuscript.

The cgf-map generates a new representation of the degrees of freedom of the system, in terms of fermions only. This is formally made possible
by a {\it non-linear transformation} of the original spin and fermionic operators,
which is easily set up working in terms of Majorana fermions, rather than 
the original operators.
The use of these kind of non-linear transformations is not new to the literature: for example it has been used in the study of the Hubbard model\cite{MatsStellanHUBBARD,KumarORIGINAL}.
In that context it has been shown how the fermionic operators representing the degrees of freedom of the conduction electrons, can be
expressed in terms of a fermionic operator (describing the holon) and three spin operators (describing the spinon).
In order to fully understand this approach, it is appropriate to review and discuss some
known properties of the fermionic operators and their representation in terms of Majoaranas.
This is done in Appendix \ref{subsec:kumar}. We invite the reader who is not familiar with these topics and in particular with the holon-spinon representation
to examine the appendix, in order to get more insight on the transformation realized by the
cgf-map that we are going to introduce in the next section.

\section{The canonical cgf-map}
\label{sec:cgfmap}

In light of the previous paragraphs and of the informations contained in Appendix \ref{subsec:kumar},
it becomes possible to give a better interpretation of the Majorana map that generates Eq.~(\ref{MajoranaHamiltonian}). The only
Majoranas that appear in $H_M$ are the three coming from the original conduction $c$-electron $\gamma_1,\gamma_2,\gamma_3$ and the
three coming from the frozen $f$-electron $\mu_1,\mu_2,\mu_3$
(to avoid cluttering of the notation we suppress the local index $\mathbf r_n$ and we refer always to the local Hilbert space if not specified otherwise).
The creation of the spinful $c$-electron is given by the spinor operator
\be\label{eq:spinorSU2}
c^\dd_c=\begin{pmatrix}
c^\dd_{c,\ua}\\
\\
c^\dd_{c,\da}
\end{pmatrix}
=
\begin{pmatrix}
\frac{\gamma_1+i\gamma_2}{\sqrt{2}}\\
\\
\frac{-\gamma_3+i(2i\mu_1\mu_2\mu_3)}{\sqrt{2}}
\end{pmatrix}.
\ee
Therefore it is immediate to identify (up to a $-\pi/2$ irrelevant phase factor) the second component of the spinor operator (\ref{eq:spinorSU2})
as the creation operator of the holon associated to a (hyper-)spinful particle described by the operator $s^\dd$:
\be\label{eq:superparticle}
s^\dd=
\begin{pmatrix}
\frac{\mu_1+i\mu_2}{\sqrt{2}}\\
\\
\frac{-\mu_3+i\gamma_3}{\sqrt{2}}
\end{pmatrix}.
\ee
It is then clear that it becomes possible to associate to each local quantum state of the
Kondo lattice model the quantum numbers of a single fermionic particle characterized by a
generalized spin, generated by an intrinsic symmetry group different from SU(2).
From this point of view the involved structure given by (\ref{eq:spinorSU2}) and (\ref{eq:superparticle})
can be easily understood, characterized and generalized. Such classification is
irrelevant for the present work, so we will leave this for future discussion\cite{MBzNilssonPREP}.


Instead we will take a much easier and straightforward direction in the following, considering the components
of this higher-dimensional spinor as independent spinless particles, in the same fashion as the usual separation
of the spinor (\ref{eq:spinorSU2}) in components $c^\dd_{c,\ua}$ and $c^\dd_{c,\da}$. In practice we consider
three spinless fermions: one for each independent component of the previous spinor, i.e. one for
the first (up) component of (\ref{eq:spinorSU2}) and two for (\ref{eq:superparticle}).

We name these three spinless fermions after the definition of their creation/annihilation operators:
\be\label{eq:cgf}
c^\dd=\frac{\gamma_1+i\gamma_2}{\sqrt{2}}, \,\,
g^\dd=\frac{\gamma_3+i\mu_3}{\sqrt{2}},\,\,
f^\dd=\frac{\mu_1+i\mu_2}{\sqrt{2}}.
\ee
For future convenience, with respect to (\ref{eq:superparticle}), we have added an extra $-\pi/2$ phase factor in the definition of $g^\dd$.

Having defined these operators, we need only to prove that the local Fock space on which they act is
(isomorphic to) the local Hilbert space of the Kondo lattice model. This will
also help us to understand the physical properties and meanings of these three particles.
The creation operators (\ref{eq:cgf}) acting on their vacuum state $|0_{cgf}\rangle$ generate an 8-dimensional
Fock space (note that the anticommutative relations between the operators
is assured by the Clifford algebra structure of the Majorana operators), which has the same dimension as the local
Hilbert space of the Kondo lattice model. As the three creation (annihilation) operators are
expressible in terms of the original electron and impurity-spin operators, it is clear that {\it if} $|0_{cgf}\rangle$
belongs to the Kondo local Hilbert space, {\it then} the other states will also belong to it.
A simple calculation shows that:
\be
c^\dd&=&c^\dd_{c,\ua} , \label{eq:c-cgfmap}\\
g^\dd&=&-\frac{1}{2}\left(c^\dd_{c,\da}+c_{c,\da}+(c^\dd_{c,\da}-c_{c,\da})2S^z_f\right), \label{eq:g-cgfmap}\\
f^\dd&=&-i(c_{c,\da}-c^\dd_{c,\da})S_f^+ \label{eq:f-cgfmap}.
\ee
We remind the reader that in our model $S_f^2=3/4$.

Looking in the local Kondo Hilbert space for the state $|0_{cgf}\rangle$, such that $c|0_{cgf}\rangle=g|0_{cgf}\rangle=f|0_{cgf}\rangle=0$, it is easy to show that
\be
|0_{cgf}\rangle=\left(c^\dd_{c,\da}|0\rangle_c\right)\otimes| \Downarrow\rangle_f =|\da\Downarrow\rangle,
\ee
where $|0\rangle_c$ is the vacuum of the original conduction electrons $c_{c,\da} |0\rangle_c=c_{c,\ua} |0\rangle_c=0$, and $| \Downarrow\rangle_f$ is the spin-down state of the original
impurity spin $S_f^-| \Downarrow\rangle_f=0$. The relations between the other states follow naturally.
We report the complete structure of the map in Tab.~\ref{tab:cgfmap}, that together with the formulas (\ref{eq:c-cgfmap})-(\ref{eq:f-cgfmap}) represents the core of our work: the $cgf$-map.

\begin{table}
\caption{\label{tab:cgfmap}States of the local Kondo Hilbert space expressed in terms of the original electron-spin quantum numbers, {\it left},
and the corresponding state in the cgf-representation, {\it right}.
The phase factors could be easily cancelled, reabsorbing them into the definitions of the different operators, but these definitions are kept for future convenience and
to maintain continuity in the notation of Appendix~\ref{subsec:kumar} and with the literature.}
\begin{ruledtabular}
\begin{tabular}{rcr}
\qquad$-|\Downarrow\rangle $&  $\longleftrightarrow$ & $g^\dd|0_{cgf}\rangle$\qquad \,\\
\qquad $-i |\Uparrow\rangle$&  $\longleftrightarrow$ & $f^\dd|0_{cgf}\rangle$\qquad \,\\
\qquad $|\da \Downarrow\rangle$& $\longleftrightarrow$ & $|0_{cgf}\rangle$\qquad \,\\
\qquad $-|\ua \Downarrow\rangle$& $\longleftrightarrow$ & $c^\dd g^\dd|0_{cgf}\rangle$\qquad \,\\
\qquad $i |\da \Uparrow\rangle$& $\longleftrightarrow$ & $g^\dd f^\dd|0_{cgf}\rangle$\qquad \,\\
\qquad $-i |\ua\Uparrow\rangle$& $\longleftrightarrow$ & $c^\dd f^\dd|0_{cgf}\rangle$\qquad \,\\
\qquad $|\ua\da\Downarrow\rangle$& $\longleftrightarrow$ & $c^\dd|0_{cgf}\rangle$\qquad \,\\
\qquad $i |\ua\da\Uparrow\rangle$& $\longleftrightarrow$ & $ c^\dd g^\dd f^\dd|0_{cgf}\rangle$\qquad \,\\
\end{tabular}
\end{ruledtabular}
\end{table}


Clearly, in principle, it is not necessary to take the Majorana approach to generate the $cgf$-map.
However  it seems improbable that a different path for the derivation could be followed,
considering the complicated structure of the fermionic operators generated.
The power of the analysis in terms of Majoranas stems from the relative easiness of the generation
of involved transformations that mix both fermion and spin operators.

Now that the spinless fermions $c$, $g$ and $f$ have been introduced, it is possible to represent the Hamiltonian (\ref{KondoHamiltonian}) making use of the $cgf$-map.
A direct calculation, that is very much simplified starting from (\ref{MajoranaHamiltonian}), leads to
\be\label{eq:cgfkondohamiltonian}
H_{cgf}=H_c + H_{de}+H_{J} + H_{\text{chem}}.
\ee
In which:
\be
H_c=-t\sum_{n,\delta} \left(c^\dd \tilde c+\tilde c^\dd c\right),
\ee
\be\label{eq:ghopping}
H_{de}&=&+t\sum_{n,\delta} \Bigl(\frac{1}{2}-f^\dd f \Bigl)\Bigl(g^\dd -g\Bigl)\Bigl(\tilde g^\dd+\tilde g\Bigl)+\\
&&\qquad\qquad-\Bigl(\frac{1}{2}-\tilde f^\dd \tilde f \Bigl)\Bigl(g^\dd +g\Bigl)\Bigl(\tilde g^\dd-\tilde g\Bigl)\nonumber,
\ee
\be
H_{J}&=&\frac{J}{4}\sum_{n} \left(1-c^\dd c-f^\dd f-g^\dd g+2c^\dd c f^\dd f\right)+\\
&&\qquad\qquad+\frac{J}{4}\sum_{n} 2g^\dd g\left\{i(c^\dd f-f^\dd c)\right\}. \nonumber
\ee
The last term $H_{\text{chem}}$ is given by the chemical potential term:
\be\label{eq:chempotcgf}
H_{\text{chem}}=-\mu^* \sum_{n}\left( c^\dd c -f^\dd f-g^\dd g+2f^\dd f g^\dd g+1\right).\nonumber
\ee
In all the previous equations we have used the same conventions of (\ref{MajoranaHamiltonian}) with
the prescription that a generic operator $\alpha$ without the tilde stands for $\alpha(\mathbf r_n)$,
while $\tilde \alpha$ represents $\alpha(\mathbf r_n+\boldgreek{\delta})$. 
The term $H_{de}$, which is the most unusual one in the Hamiltonian (\ref{eq:cgfkondohamiltonian}),
describes a density-correlated hopping for the $g$-fermions and is responsible for
the description of the double exchange mechanism. 
It is the appearance of this term that makes this three-fermion representation of the Kondo lattice very successful in the
description of ferromagnetism, as will be made clear by the mean-field analysis in the next section.
We stress that these kinds of non-trivial hopping structures
are a typical consequence of the non-linear transformations of the type (\ref{eq:spinorSU2}).
Similar situations are encountered for example in both the context of the one-band Hubbard model\cite{MatsStellanHUBBARD,KumarORIGINAL}
and the t-J model\cite{Hellberg:1991mq,*Hellberg:1991wm}.


It is evident that the Hamiltonian (\ref{eq:cgfkondohamiltonian}), the mapping of
Tab.~\ref{tab:cgfmap} and the equations (\ref{eq:c-cgfmap})-(\ref{eq:f-cgfmap}), can be demonstrated by {\it direct inspection},
without passing through the procedure that produces $H_M$. In fact
one can simply consider the Hamiltonian (\ref{eq:cgfkondohamiltonian}), i.e. the $cgf$-form of (\ref{MajoranaHamiltonian}), as an ansatz,
and then using formulae (\ref{eq:c-cgfmap}), (\ref{eq:g-cgfmap}) and (\ref{eq:f-cgfmap}) the usual Hamiltonian (\ref{KondoHamiltonian}) of the KLM is recovered.
Hence, using the very general definitions of Eq. (\ref{eq:cgf}), the $cgf$-map generates a different, indirect and alternative demonstration of the faithful Majorana representation
of the Kondo lattice model given by equations (\ref{MajoranaHamiltonian}) and (\ref{eq:spinorSU2}).

To understand the meaning of the three fermions (\ref{eq:cgf}) it is useful to express some physically interpretable operators in terms of them. The easiest expressions are given for
the spin-up conduction electron density and the impurity $\hat z$-oriented spin operator:
\be\label{eq:cgfeasydensities}
c^\dd_{c,\ua} c_{c,\ua} =c^\dd c,\qquad S_f^z=f^\dd f-\frac{1}{2}.
\ee
Of much greater interest is the density operator of the spin-down component of the conduction electron:
\be\label{eq:cgfcdowdensity}
c^\dd_{c,\da} c_{c,\da} = 1- f^\dd f-g^\dd g+2f^\dd fg^\dd g.
\ee
This operator, quadratic in the original representation, becomes partly quartic if represented on the $cgf$-operators. This is not surprising, considering that the map
is built on the idea that the $c_{c,\da}$-electron must be interpreted as the holon of (\ref{eq:superparticle}). Given that two particles contribute to the
constitution of the $c_{c,\da}$-fermion, it must happen that their densities sum up properly.
In a more concrete fashion, it is possible to think that the electric charge density associated with
the fermion degree of freedom $c_{c,\da}$ (i.e., the down component of the physical electron mode of charge $e=1$),
decomposes into two channels given by the two primitive particles of which it consists.
The non-quadratic form of the chemical potential term $H_{\text{chem}}$ is a direct consequence of (\ref{eq:cgfcdowdensity}).
The coefficient $\mu^*$ will determine not only the
amount of total electric charge density (as it does in the usual linear case), but it will affect also how the density of $c_{c,\downarrow}$ electrons
is redistributed between the two channels $g$ and $f$.

Although unconventional this technique of fermion decomposition is not a complete novelty in the literature. Similar
approaches have been followed for example in the study of the t-J model\cite{Hellberg:1991mq,Hellberg:1991wm}
and in a quite general fashion we can classify them into the framework of generalized Bogoliubov transformations.

Substituting (\ref{eq:cgfeasydensities}) into (\ref{eq:cgfcdowdensity}), we can rewrite
\be
c^\dd_{c,\da} c_{c,\da} = \frac{1}{2}+\left(g^\dd g- \frac{1}{2}\right)2S_f^z ,
\ee
or
\be
g^\dd g=\left(c^\dd_{c,\da} c_{c,\da} - \frac{1}{2}\right)2S^z_f+\frac{1}{2}.
\ee
The latter equation displays the nature of the $g$-fermion density: it is generated
by the original $c_{c,\da}$-density and an impurity-spin dependent particle-hole transformation $c_{c,\da}\leftrightarrow c^\dd_{c,\da}$.
Schematically we have on each generic local state $|\alpha\rangle$
\be
&\text{if}&\quad S_f^z|\alpha\rangle=\frac{1}{2}|\alpha\rangle\quad\quad c^\dd_{c,\da} c_{c,\da}= g^\dd g,\\
&\text{if}&\quad S_f^z|\alpha\rangle=-\frac{1}{2}|\alpha\rangle\quad c^\dd_{c,\da} c_{c,\da}= 1-g^\dd g.
\ee
This means that the density operator $g^\dd g$ counts the number of down-spin conduction electrons (holes)
on the sites where the local impurity points up (down). 

It is evident that the $g$ and $f$ fermions represent very non trivial spin-electron excitations, whose nature will be made understandable by our mean-field analysis.
However in this section it is appropriate to point out an intriguing parallelism between our $f$ fermion and the composite fermion used
in the large-N approximation. In large-N studies of the Kondo lattice
an auxiliary fermion operator is used to represent the local spins\cite{Coleman:qf,Coleman:1983ys}.
Because of the Kondo interaction this auxiliary fermion develops dynamics and becomes the most intriguing
excitation of the system: the heavy fermion. This heavy fermion is shown to have a composite nature: it is a bound state of the local
spin and the conduction electron. In particular it binds the creation of a conduction electron to a spin-flip
of the impurity-spin on the same site.
It is evident from Eq.~(\ref{eq:f-cgfmap})
that our $f$ fermion is very similar to this large-N composite fermion.
There are of course some differences (in particular the particle-hole linear combination of the conduction electrons),
but this parallelism of our formalism with the more known and quite successful large-N approximation is very interesting.



\section{Mean-field analysis}
\label{sec:HartreeFock}

As a first approach to the Hamiltonian (\ref{eq:cgfkondohamiltonian}) we perform a mean-field study,
to explore the possible ground states and understand
the nature of the degrees of freedom that we are using to describe the system.
To follow this path the symmetries of $H_{cgf}$ must be identified.
The analysis of the symmetries 
can be easily done also if the Hamiltonian is written down in the Majorana representation (\ref{MajoranaHamiltonian}),
looking for the operators that commute with it. 
Among them we identify the non-trivial operator
\be
A_3(n)&=&-i\gamma_1(n)\gamma_2(n)-i\mu_1(n)\mu_2(n)\nonumber\\
&=&c^\dd(n)c(n)+f^\dd(n)f(n)-1.\label{eq:PruschkeSYMM}
\ee
It is straightforward to check that\footnote{The operator $A_3=\sum_n A_3(n)$ commutes with the Hamiltonian for any value of the chemical potential $\mu^*$. For the particular value $\mu^*=0$ also two
other operators $A_1$ and $A_2$, with form similar to $A_3$, can be found.}
\be\label{eq:symmetry}
\left[H,\sum_n A_3(n)\right]=0.
\ee

In our mean-field analysis we decided to enforce the symmetry (\ref{eq:symmetry}), so we imposed the commutation
between the mean-field $cgf$-Hamiltonian $H^{MF}_{cgf}$ and the operator $\sum_n A_3(n)$. The rationale behind this
choice is that the breaking of this symmetry is a necessary condition for superconductivity that we do not wish to include in the study.
In fact the consequences of our choice are:
\be
&&\langle c^\dd f^\dd\rangle=\langle g^\dd c^\dd\rangle=\langle g^\dd f^\dd\rangle=0,\nonumber\\
&&\langle g^\dd f\rangle=\langle g^\dd c\rangle=0,\label{eq:conditionsSYMMETRY}\\
&&\langle f^\dd \tilde g^\dd\rangle= \langle f^\dd \tilde g\rangle=\langle \tilde f^\dd g^\dd \rangle=\langle \tilde f^\dd g\rangle=0.\nonumber
\ee
This means that there exist no open channel for the hybridization of the $g$-fermions with $c$- and $f$-fermions, so the pairing
order parameter $\langle c_{c,\ua} c_{c,\da} \rangle$ will always be zero, as can be checked using the formula
$$
c_{c,\da}=-g^\dagger-f^\dagger f (g-g^\dagger),
$$
and making use of Wick's theorem to decompose the higher order operators.

The enforcement of this symmetry hides some very subtle and unexpected surprises.
In fact, rewriting (\ref{eq:PruschkeSYMM}) in terms of the usual impurity-spin and conduction-electron operators, we obtain:
\be
A_3&=&\frac{1}{2}\Big\{ 2S_f^z+(c^\dd_{c,\ua} c_{c,\ua} +c^\dd_{c,\da} c_{c,\da})+\\
&&\qquad\qquad+(c^\dd_{c,\ua} c_{c,\ua} -c^\dd_{c,\da} c_{c,\da})-1  \Big\}\nonumber\\
&=&\frac{1}{2}\left( 2S_f^z+n_c+2m-1\right),\nonumber
\ee
where the site index $n$ has been suppressed to not clutter the notation.
This implies the existence of a very non-trivial relation between the electron magnetization $\langle m\rangle=\langle c^\dd_{c,\ua} c_{c,\ua} -c^\dd_{c,\da} c_{c,\da}\rangle/2$,
the electron density $\langle n_c\rangle =\langle c^\dd_{c,\ua} c_{c,\ua} +c^\dd_{c,\da} c_{c,\da}\rangle$ and the spin polarization $\langle S_f^z\rangle$.
The existence of this non-trivial relation
has been recently discovered numerically making use of DMFT+NRG methods\cite{Peters:2012rq} in infinite dimensions and DMRG\cite{Peters:2012fc}
in the one-dimensional case. To be consistent with the literature, we call the $A_3$ operator the ``commensurability operator''.
The sign differences between our definition of commensurability and the one that can be found in the numerical studies\cite{Peters:2012rq,Peters:2012fc}
is due to an opposite convention for the up direction of the spin polarization axis.


Therefore the first concrete result of our analysis is the identification of this symmetry, that furnishes a theoretical justification to
the ``commensurability'' parameter. We want to remark that we have not yet restricted our analysis to the one-dimensional case,
but we are still dealing with an arbitrary number of dimensions, therefore these conclusions refer to both the DMRG\cite{Peters:2012fc}
and the DMFT+NRG results\cite{Peters:2012rq}.

It is useful for the discussion to write down explicitly the form of $H^{MF}_{cgf}$ obtained enforcing the symmetry (\ref{eq:symmetry}):
\be\label{eq:MFHamiltonian}
H^{MF}_{cgf}=H^{MF}_{cf}+H^{MF}_g+H^{MF}_{\text{chem}}+H^{MF}_{\text{shift}},
\ee
where
\begin{widetext}
\be
H^{MF}_{cf}&=&-t\sum_{n,\delta} \left(c^\dd \tilde c+\tilde c^\dd c \right)+ t\sum_{n,\delta} \left(\frac{1}{2}-f^\dd f \right)\mathcal S_n-\left(\frac{1}{2}-\tilde f^\dd \tilde f \right) \mathcal P_n+\\
&&\quad\quad+\frac{J}{4}\sum_{n} 2 \left(\mathcal G_n+ \mathcal I_n\right)\left\{i(c^\dd f-f^\dd c)\right\}-2\mathcal R_n (c^\dd f+f^\dd c)+\sum_n (2 \mathcal C_n -1)f^\dd f+(2 \mathcal F_n-1) c^\dd c,\nonumber
\ee
\be\label{eq:HmfG}
H^{MF}_g&=&t\sum_{n,\delta} \left(\frac{1}{2}-\mathcal F_n \right)\Bigl(g^\dd -g\Bigl)\Bigl(\tilde g^\dd+\tilde g\Bigl)-\left(\frac{1}{2}-\mathcal F_{n+1}\right)\Bigl(g^\dd +g\Bigl)\Bigl(\tilde g^\dd-\tilde g\Bigl)+\frac{J}{4}\sum_{n}(-1-4  \mathcal I_n)g^\dd g,
\ee
\be
H^{MF}_{\text{chem}}=-\mu^* \sum_n  \Bigl(c^\dd c -f^\dd f+2\mathcal G_n f^\dd f +2\mathcal F_n g^\dd g +1\Bigl),
\ee
\end{widetext}
and $H^{MF}_{\text{shift}}$ represents the total, mean-field dependent, shift of the energy. The definitions for the various mean-fields are:
\be\label{eq:meanfieldsG}
S_n=\Bigl\langle \Bigl(g^\dd -g\Bigl)\Bigl(\tilde g^\dd +\tilde g \Bigl)\Bigl\rangle,&\quad& P_n=\Bigl\langle \Bigl(g^\dd+g\Bigl)\Bigl(\tilde g^\dd-\tilde g \Bigl)\Bigl\rangle,\nonumber\\
\nonumber
\ee
\be
\mathcal I_n=-\frac{i}{2}\Bigl(\langle c^\dd f\rangle-\langle f^\dd c\rangle\Bigl),&\quad& \mathcal R_n=\frac{1}{2}\Bigl(\langle c^\dd f\rangle+\langle f^\dd c\rangle\Bigl), \nonumber
\ee
\be\label{eq:meanfieldsCF}
\mathcal F_n=\langle f^\dd f\rangle, \quad \mathcal C_n=\langle c^\dd c\rangle, \quad \mathcal G_n=\langle g^\dd g\rangle.
\ee
With the subscript $n$ we want to remark the fact that modulation of the mean-fields are allowed in general.
Of course modulations with wave-vectors different from $\mathbf K=0$ imply the study of larger unit cells.


The $H^{MF}_{cf}$ describes a delocalized spinless fermion (electron) that hybridizes with a lattice of $f$-impurities; while the $H^{MF}_{g}$ is
the fermionic representation of the generalized transverse field Ising model\cite{SachdevBOOK,KitaevBOOK}. The exact dynamics of the two subsystems depend
on the specific structure of the mean-fields.

If $J=0$, the $c\,$-$f$ hybridization does not
takes place, so the $f$ fermions give rise to a flat band, while the $c$-fermions produce the usual multidimensional cosine free-band. When $J$ is increased
the two species hybridize, causing the opening
of a gap in the band structure, with the typical {\it avoided crossings}.

Although the $H^{MF}_g$ has an unconventional form, it is important to
stress that according to (\ref{eq:cgfeasydensities}) one obtains
\be
\mathcal F_n=\langle f^\dd (n) f(n)\rangle=1/2-\langle S_f^z(n)\rangle,
\ee
therefore if the impurity-spin order
parameter is constant in space (ferromagnetic order), then the $g$-fermions are described by a simple non-interacting model, while any space modulation introduces
p-wave pairing terms in the $g$-Hamiltonian.

From these considerations it is evident that the structure of the $\mathcal F_n$ field plays a central role in the mean-field Hamiltonian.
In fact it is the scattering of the $g$-electrons on the modulations of $\mathcal F_n$ that causes the opening of a gap in the $g$-band.
Moreover one should keep in mind that the magnitude of $\mathcal F_n$ also determines the bandwidth of the $g$-fermion band.
This band-renormalization is the most unconventional feature
of the Hamiltonian (\ref{eq:MFHamiltonian}) and it is remarkable, especially considering that we obtain it in a mean-field framework.
Indeed band renormalization effects do not typically appear in mean-field contexts, while they are obtained in more involved
approximation schemes, such as Gutzwiller projection methods.

\begin{figure}[t!]
\includegraphics[width=1\columnwidth]{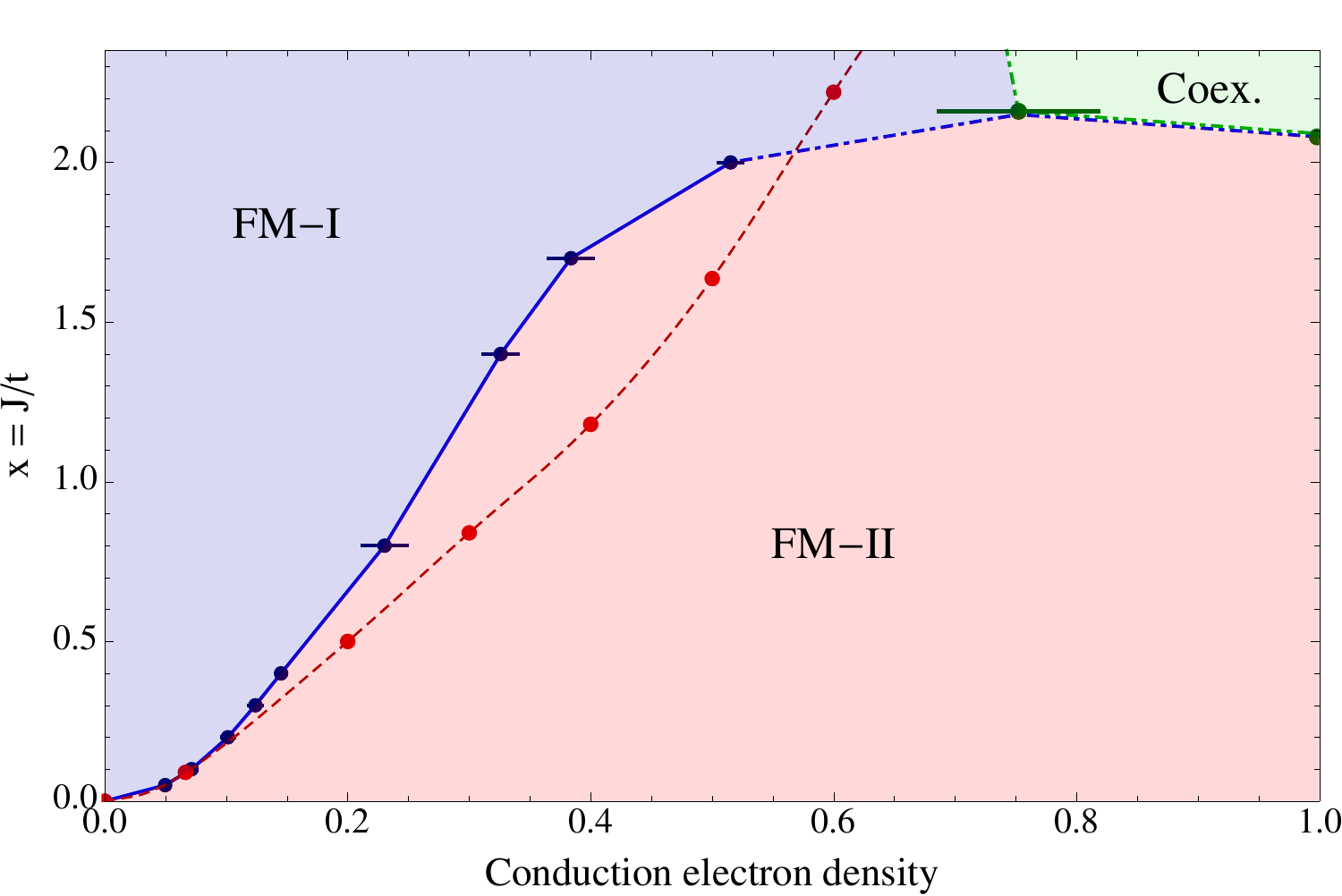}
\caption{The $x=J/t$ vs $n_c$ phase diagram. The thick continuos blue line represents the phase transition line between the FM-I and FM-II phase
and should be compared with the dashed red line, that shows the ferromagnetic-paramagnetic transition characterized by the DMRG calculations in
Ref. \onlinecite{McCulloch:2002ss} and \onlinecite{Peters:2012fc}. The dashed blue line and the dot dashed green line represents the transition
line between the ferromagnetic phases and the region of phase coexistence between the KI phase and the FM-I phase. The point at $x=0$ is an extrapolation
of the results, because our approach cannot be applied for that specific coupling value.
\label{fig:phase-diag-low}}
\end{figure}



From all these considerations it is clear that it is very difficult to predict the properties of the mean-field ground state, so that the only way to tackle the problem is numerically.
We did it following the algorithm presented in Appendix \ref{app:A}, generating a system of non linear equations
whose solutions coincide with the values of the order parameters for the mean-field solutions. In order to find all the possible solutions of the
non linear system, we created a grid in the entire parameter space and used each point of the grid as starting point for a Newton-Raphson root finder.
We repeated this procedure for different values of chemical potential and Kondo coupling, mapping the full phase diagram\footnote{We accepted only the solutions
that were able to solve the non linear system within a better accuracy than $10^{-12}$.}.
This procedure permitted us to find not only the mean-field ground state, but also higher energy mean-field states and to follow them in their
complicated evolution in the phase diagram.

In the following we will focus our attention on translationally invariant solutions, which 
should be dominant away from half-filling and from the weak coupling limit;
therefore we will consider constant values for all the mean-fields on the entire lattice.
We must stress the fact that this does not mean that we are forcing the system into a ferromagnetic state. In fact paramagnetic
states, which posses this kind of translational invariant property, will still be in principle included into the sector of the theory that we
are going to analyze. What will be excluded are states and effects that are characterized by non-local correlations between different sites.
So we expect our solution to not be able to capture the physics of the RKKY liquid phase, for example.
To describe (at the mean-field level) states who posses these kind of non-local correlations,
we should allow for the spatial modulation of the mean-fields.
Only at the end will we discuss this option, considering the possibility of adding a staggered modulation for the mean-fields, restricting the analysis
to the half-filled system. In this way it would be possible to see how the RKKY effect enters
into the Hamiltonian and why antiferromagnetic or spiral orders can appear at mean-field level.
Unfortunately the analysis will reveal that a neat separation between the RKKY and Kondo effect
is in general impossible. To go beyond the known results\cite{Fazekas:1991kc} we should allow
for the competition of these two effects, but this would make the mean-field analysis more involved.
Therefore, in this work, we do not consider this extremely general case that in our opinion should
instead be tackled with different approximation schemes\cite{ChristmaMajorana}. However we will point out the
connection with the known mean-field treatments of the RKKY effect in the KLM, and provide
a discussion of the half-filled system showing how the paramagnetic Kondo Insulating phase becomes the mean filed ground state at high coupling.
Since this phase is also translational invariant our mean-field study becomes quite meaningful and, in particular, very efficient in the analysis of the
quasiparticle gap.


Although our mean-field algorithm works for any temperature $T$ and number of dimensions,
we performed a study only in one dimension and in the $T\rightarrow0$ limit;
we leave the other cases for future studies.


\subsection{Ferromagnetic solutions away from half-filling}\label{subsec:TIS}
We will now focus our attention on the results that we obtained imposing translational invariance, i.e.
assuming a constant value of all mean-fields on the entire lattice.
The translational invariant Hamiltonian, that we indicate as $H^{MFTI}_{cgf}$, looks like:
\begin{widetext}
\be\label{eq:MFHamiltonianTI}
H^{MFTI}_{cgf}=&-&t\sum_{n} c^\dd \tilde c+\tilde c^\dd c-4t (\mathcal S-\mathcal P) \sum_{n} f^\dd f +
\frac{J}{4}\sum_n 2 \left(\mathcal G+ \mathcal I\right)\left\{i(c^\dd f-f^\dd c)\right\}+(2 \mathcal C -1)f^\dd f+(2\mathcal F-1) c^\dd c+ \nonumber\\
&+&2t\left(\frac{1}{2}-\mathcal F \right)\sum_{n} g^\dd \tilde g+\tilde g^\dd g +\frac{J}{4}\sum_n(-1-4  \mathcal I)g^\dd g+H^{MF}_{\text{chem}}+H^{MF}_{\text{shift}},
\ee
\end{widetext}
where we put $\mathcal R=0$, without loosing generality and
en passant we note that $\mathcal S-\mathcal P=\langle g^\dd \tilde g+\tilde g^\dd g\rangle$.

The $g$-fermion subsystem is described by a trivial non-interacting Hamiltonian. Therefore all the mean-fields $\mathcal G$ and
$\mathcal S-\mathcal  P$ may be computed analytically, as functions of the other variables.

We analyzed the system for discrete values of the adimensional coupling parameter $x=J/t$ between $0.05$ and $6$ and
for different values of the chemical potential, in order to have a description of the most relevant region of the ($x$-$n_c$) phase diagram.
For each value of $x$ and $\mu$ the non-linear system was solved and the free energy $E-TS-\mu^* N$
was used to order the different solutions and to identify the mean field candidate ground-state.
In general the final picture that we obtain can be split in four regions: the first at half-filling and the other three away from half filling, respectively at
low coupling ($x\lesssim2$), intermediate coupling ($2\lesssim x\lesssim3$) and high coupling ($x>3$). In the second
region it is possible to obtain, from our mean field analysis, quantitative information about the structure of the ground state
for any value of the filling. Instead in the last two regions our approach becomes less efficient, providing only indications
on the nature of the system.


In the low coupling regime we discovered the existence of two ferromagnetic phases FM-I and FM-II, divided by a second order phase transition that takes place at
the critical conduction electron density $n^F_{crit}(J)$.
This part of the phase diagram is plotted in Fig.~\ref{fig:phase-diag-low}, where it is possible to see how closely
our line of phase boundary is, when compared with the one found by DMRG methods\cite{Peters:2012fc,McCulloch:2002ss}.
This result improves significantly the available previous\cite{Fazekas:1991kc} mean-field ones.
In particular we do not identify the unphysical (global Kondo-singlet like) paramagnetic solution, discovered by other mean-field methods, but
we correctly find a ferromagnetic state stabilized by the Kondo effect. This is remarkable,
since we do not force the system to a magnetic sector, but just to a translational invariant one. The paramagnetic solution appears
only at half-filling, which is the only region of the phase diagram where such a solution is expected.

We invite the reader to not confuse the paramagnetism
induced by non-local correlations, with the one that we analyze. The non-local kind of paramagnetism has been excluded by our study when we chose to
not spatially modulate our mean fields. In fact the zone occupied by the paramagnetic states away from half-filling is instead covered by the FM-II state, which is a
different kind of ferromagnetic phase.

It is important to point out how the FM-II phase at low coupling is a very competitive state, that has a better energy than the
typical trial variational ground-states\cite{Fazekas:1991kc} obtained via the introduction of spiral order in the impurity spins.
This behavior marks the important role, also at low coupling, of the Kondo effect, which is able to stabilize a low energy ferromagnetic state,
if properly considered.
Since it is well proved\cite{Juozapavicius:2002ly,Gulacsi-:2004nx,Gulacsi:2006oq} that in this regime the spin-spin correlation function is peaked at $2k_F$,
it is expected that variational states with (properly modulated) spiral spin order are energetically more competitive; the existence of the FM-II phase shows that
the ordering of the spins alone is not enough to obtain a physically relevant trial ground state and the Kondo effect cannot be disregarded. Of course these considerations
are valid in the coupling regime that we considered. We did not perform any analysis of the extremely low coupling regime, where it is not excluded that the spiral ordered states
can become dominant.

\subsubsection{Ferromagnetism in the low coupling limit}\label{sec:lowcoupling}

\begin{figure}[t]
\includegraphics[width=1\columnwidth]{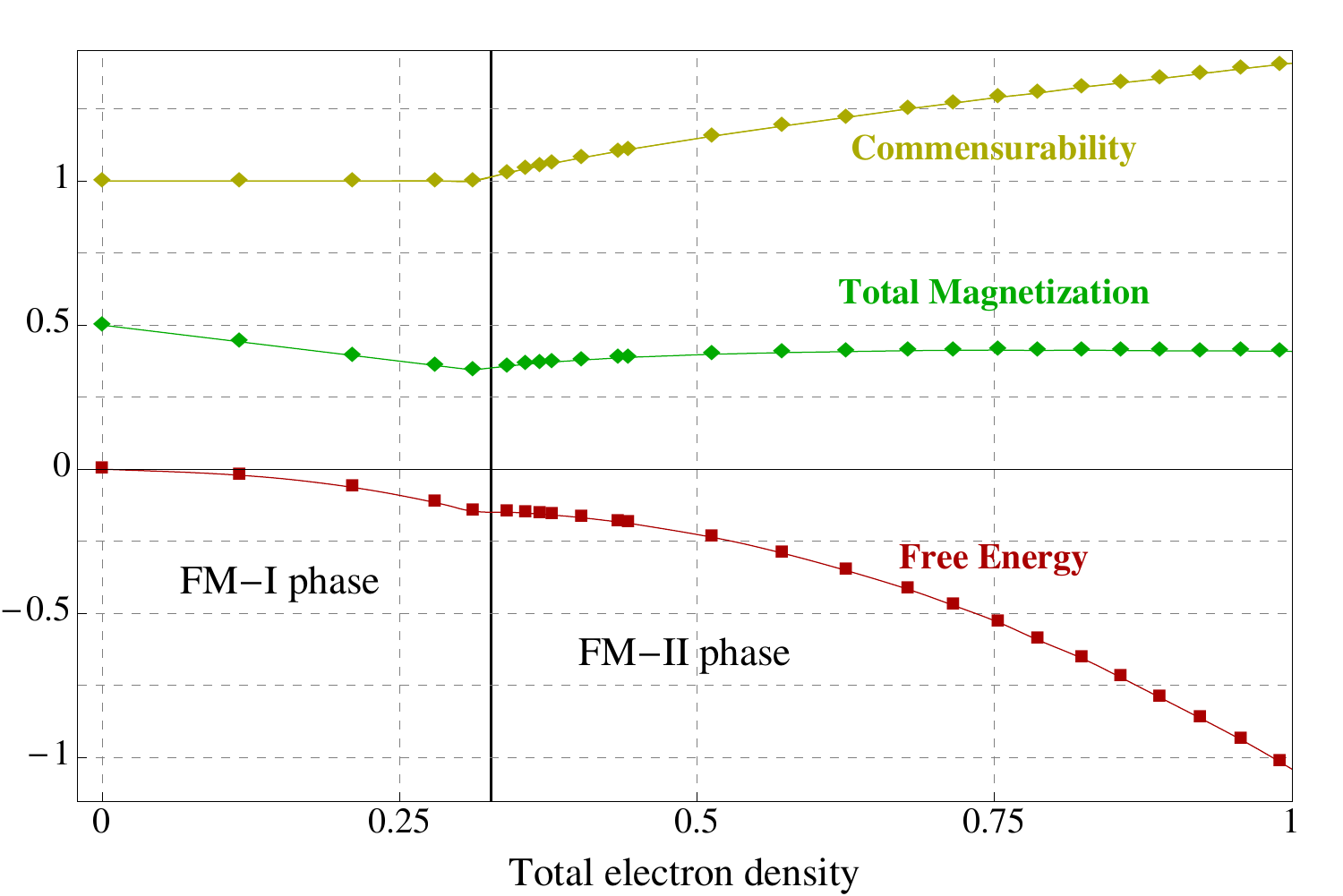}
\caption{Free energy {\it red} (plotted in units of $4t/\pi$), total magnetization $\langle S_f \rangle+\langle c_\ua^\dd c_\ua -c_\da^\dd c_\da \rangle/2$ {\it green}
and commensurability $\langle A_3\rangle+1=\langle c^\dagger c+ f^\dagger f\rangle$ {\it yellow}, versus total electron density per site $n_c$, for the coupling $x=1.4$.
Evidently in the FM-I phase the magnetization is described by the known relation $|1-\langle\hat{n}_c\rangle|/2$.
The critical line, {\it continuos vertical}, is put in correspondence of the critical density $n_{crit}^{F} \approx 0.35$. Beyond that value the commensurability increases and
the total magnetization becomes quickly almost constant. It is evident that there exists a discontinuity in the derivate of
the free energy respect to the electron density, which may indicate a phase-transition of the Lifshitz kind, as explained in the text.
\label{fig:CME}
}
\end{figure}

\begin{figure}[t]
\includegraphics[width=1\columnwidth]{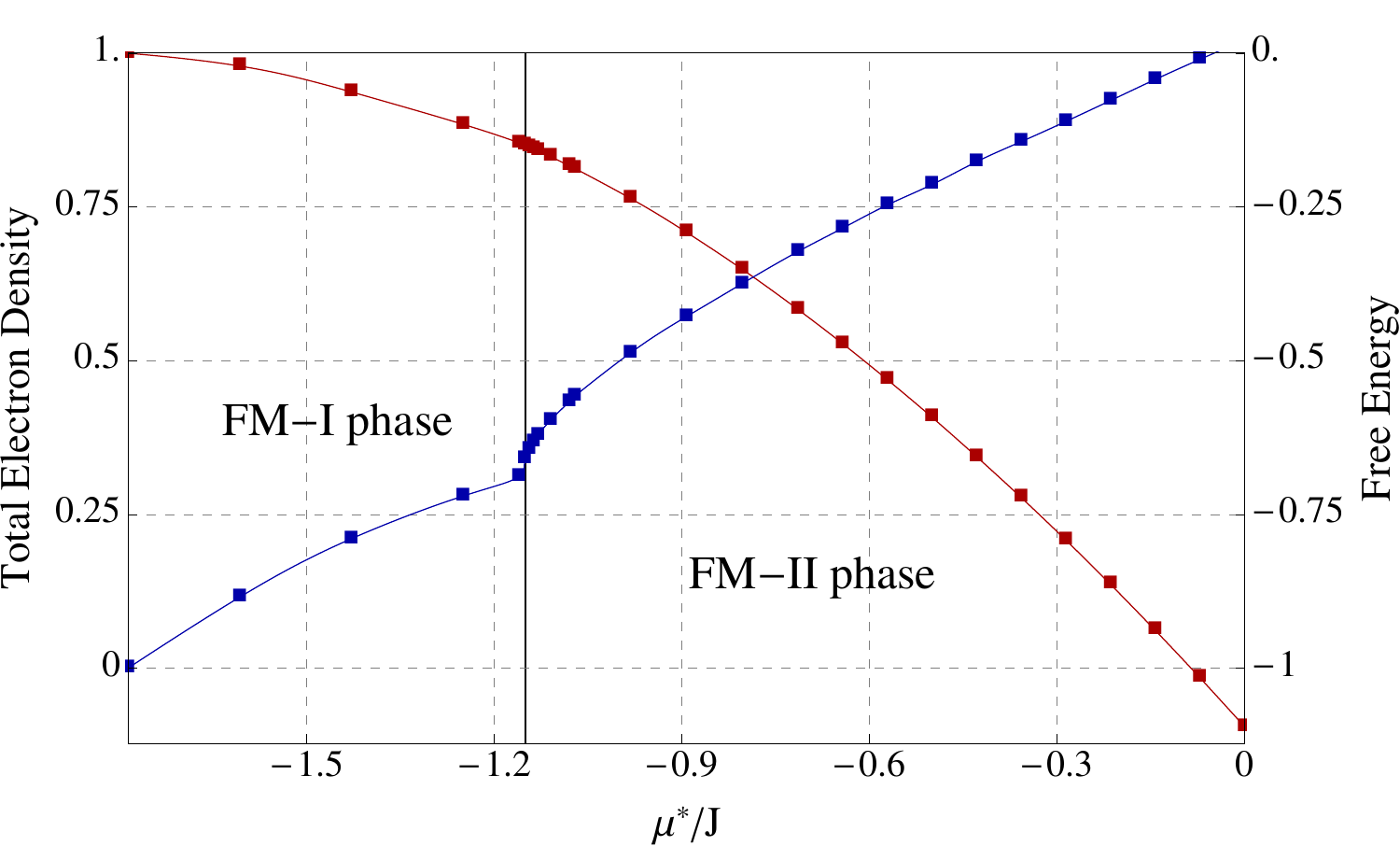}
\caption{The Free energy {\it red} (plotted in units of $4t/\pi$), and the total electron density {\it blue}, versus the (rescaled) chemical potential $\mu^*/J$ for $x=1.4$. The vertical flex
of the density line, due to the contribution of the divergent density of state of the $c$-like band is very well visible.
\label{fig:Edensity}
}
\end{figure}

\begin{figure}[t]
\includegraphics[width=1\columnwidth]{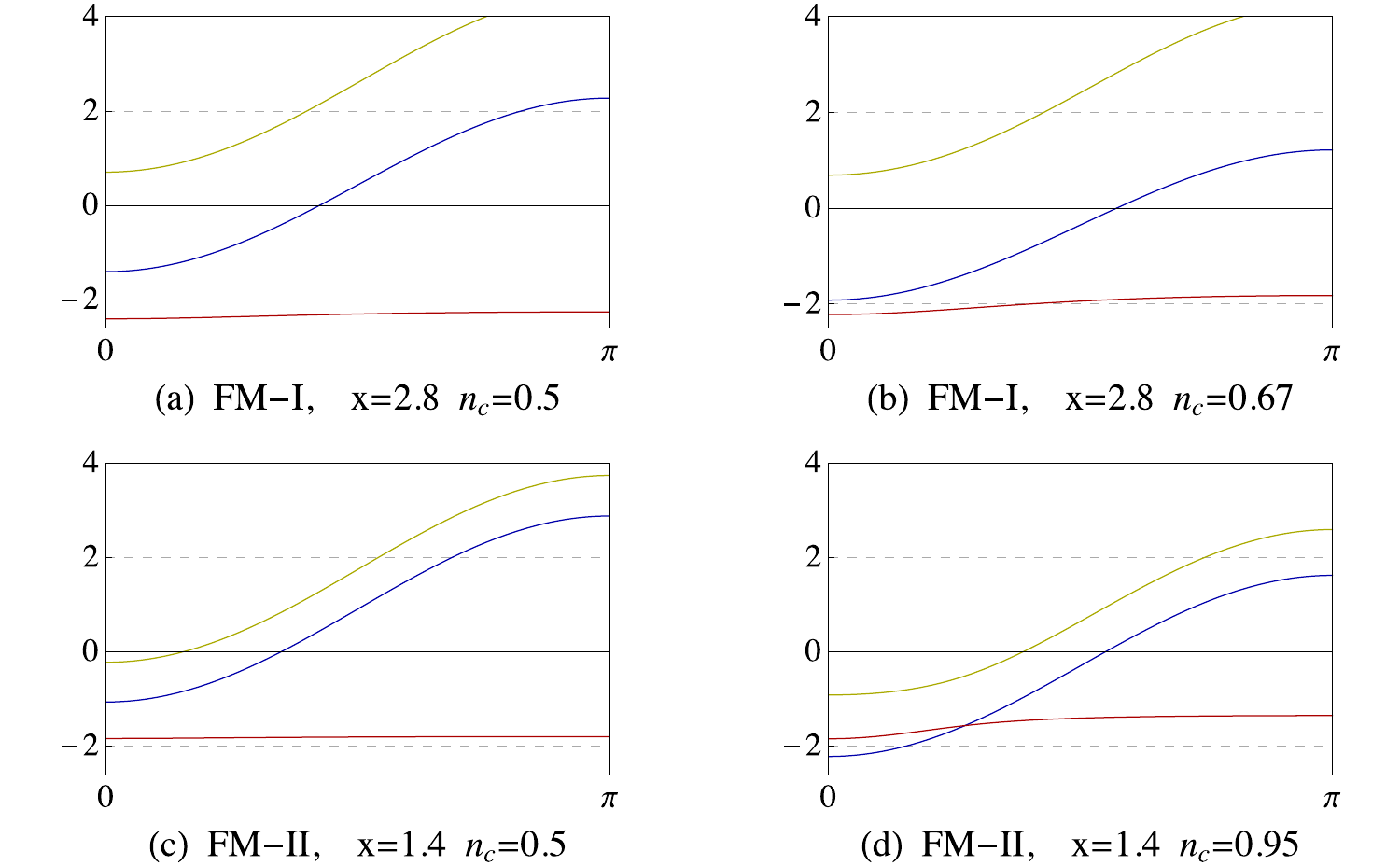}
\caption{The $cgf$ mean-field band structure for different parameter values. In (a) and (b) the FM-I band structure respectively at $x=2.8$, $n_c\approx0.5$ (i.e. $\mu^*/J=1.8$) and
$x=2.8$, $n_c\approx0.67$ (i.e. $\mu^*/J=1.49$). In (c) and (d) the FM-II phase at $x=1.4$, $n_c\approx0.5$ (i.e. $\mu^*/J=1.375$) and $x=1.4$, $n_c\approx0.95$ (i.e. $\mu^*/J=0.2$).
The $g$-band (blue) presents always a Fermi surface at zero energy; in the FM-I phase the yellow ($c$-like) band is completely empty, while in the FM-II phase it is partially filled. The
red $f$-like band is instead always filled. The hybridization gap is well visible in all the plots. The band structure is symmetric under $k\rightarrow -k$ and on the y-axis the energy is always
given in units of $t$.
\label{FIG:band28FM}
}
\end{figure}

In the coupling regime $x\lesssim 2$, we found for each value of $\mu^*$ and $x$ two possible solutions. Varying the two parameters
these two solutions formed two sets of adiabatically connected mean field states. One of the families was clearly very well separated in energy and so we discarded
it, focusing only on the lowest (free) energy mean-field states. On this branch, as mentioned previously, it is possible to identify two different phases:
the phase FM-I that extends from $n_c=0$ to $n_c=n^F_{crit}(J)$ and the phase FM-II that goes from $n_c=n^F_{crit}(J)$
to $n_c=1$. Though both the phases are ferromagnetic, they are characterized by different physical properties. In Fig.~\ref{fig:CME}, we plot
the values of the commensurability, total magnetization and Free energy, versus the conduction electron density, using as example the coupling $x=1.4$.
Evidently there exist a discontinuity in the behavior of these quantities at $n^F_{crit}\approx0.35$, that corresponds to the critical chemical potential $\mu_{crit}/J\approx-1.1$.
Such discontinuities in the derivatives of the curves continue to exist also if they are plotted respect to the chemical potential, with the exception of the Free energy.
Indeed, as shown in Fig.~\ref{fig:Edensity}, the Free energy curve looks continuous and we were not able to resolve any discontinuity in the derivate. A more detailed
analysis reveals the origin of the discontinuity in the derivate of the Free energy curve in Fig.~\ref{fig:CME}. Starting from the plots
of the mean field bands for the two phases FM-I and FM-II (some examples are plotted in Fig.~\ref{FIG:band28FM}), it is easy to understand how the system
goes from the FM-I phase to the FM-II phase, via a Lifshitz transition\cite{LifshitzTransOriginal,*Blanter:1994kxBIS}, when the
$c$-like band (curve always on the top in Fig.~\ref{FIG:band28FM}) crosses the Fermi level at zero energy. The discontinuity in
the Free energy versus the electron density is therefore due to the divergent contribution to the density of states, generated by the bottom of the $c$-like band
that gets occupied. This is also consistent with the behavior of the density versus the chemical potential, where a vertical flex in correspondence of $n^F_{crit}$ is present.

Focusing now on the FM-I phase,
we identify this state with the ferromagnetic ground state discovered by DMRG calculations. In fact, as shown for example in Fig.~\ref{fig:CME}, we recognize that
this state has a density dependent total magnetization that correctly\cite{Tsunetsugu:1997vn,Peters:2012fc}
goes linearly with the total electron density as $|1-\langle n_c\rangle |/2$,
starting from a totally ferromagnetic state at infinitesimal density\cite{Tsunetsugu:1997vn,Sigrist:1991fu}.
It is also evident from Fig.~\ref{fig:CME} (but of course this is true for any ground state of the FM-I phase) that
on the FM-I ground-state the commensurability parameter is equal to one for each value of the electron density,
exactly as in the DMRG solutions\cite{Peters:2012fc}.

Beside these quantitative agreements, we discover that also the physical picture of the ``spin-selective Kondo insulator'' (SSKI) is perfectly consistent with
the picture offered by our mean field ground state in the FM-I phase.
The physics behind the SSKI, as proposed in Ref. \onlinecite{Peters:2012fc,Peters:2012rq},
is very interesting and our $cgf$-description of the system
exposes it perfectly.
Starting from the one conduction electron limit, it is possible to understand the main features of this mechanism.
The one electron system is notoriously
ferromagnetic as can be proved analytically\cite{Tsunetsugu:1997vn,Sigrist:1991fu} or understood invoking the double exchange mechanism\cite{Gulacsi-:2004nx}.
Indeed, since the electron hopping operator preserves the spin of the electron and considering that
the system wants to maximize the energy gain from the antiferromagnetic coupling, it must happen that all the impurity spins
align in the direction opposite to the spin of the only conduction electron present.
This effect is taken into account in $H^{MFTI}_{cgf}$ by the
density-correlated $g$-hopping term (\ref{eq:ghopping}): it is clear that
in order to maximize the gain from the kinetic energy contribution the system will develop ferromagnetism. Therefore with an infinitesimal
electron density the result must be $\mathcal F=1$, so that via (\ref{eq:cgfeasydensities}) $\langle S_f^z\rangle=1/2$ and the bandwidth of the $g$-band is maximized\footnote{The
mean-field could also take value $0$, with all the spins pointing down and implying the reversing of the $g$-band.
This is not in contradiction with the results and it is a simple feature of the choice of the map. In
fact such an inversion corresponds to the simultaneous application of a parity transformation for the spin polarization axis and
a particle-hole transformation in the down-electron band. Therefore the $g$-fermions will not describe electrons propagating in an empty band,
but holes in a filled one. It is clear that the same physical picture is recovered, although it is expressed in very complicated terms. In the
$cgf$-map the choice of the direction of the impurity spins determines also a choice in how the charge of the electrons with opposite spin is counted.}.
It is easy to understand this process in the semiclassical picture, i.e. turning off
the spin-flip part of the electron-impurity spins interaction and considering only the Ising like part.

Inserting now more electrons into the system and considering the effect of the spin flip processes that the electrons experience
scattering against the impurity spins, the situation becomes more involved. The configuration where all the
electrons share the same spin is not energetically optimal nor consistent with the existence of the electron-spin scattering,
so also the band of $c_{c,\ua}$ electrons (minority-electrons) must be partially filled. This is problematic
for the system, because the up-electrons have spins parallel with the ferromagnetically ordered impurity spins (majority-spins).
To solve this problem the system binds the minority-spins (generated by the scattering of the electrons against the ferromagnetically ordered majority spins)
to the minority-electrons, performing a sort of ``effective annihilation'', via the creation of Kondo singlets.
The latter become the relevant objects of the system and the minority spins and minority electrons ceases to exists as independent degrees of freedom,
becoming only highly correlated components of the singlets.
Of course this process has to happen not just locally, but taking into account the delocalization of the singlets on the entire system.
Obviously the wave-function of the quantum liquid formed by the delocalized Kondo-singlets must be entangled with the
one describing the Fermi liquid of the majority electrons, because also (part) of the $c_{c,\da}$ must participate in the creation of the singlets.

These singlets are responsible for
the formation\cite{Peters:2012fc,Peters:2012rq} of the SSKI that is described by the two $c\,$-$f$ hybridized bands
(in the following called $c$-like and $f$-like bands) in the Fig.~\ref{FIG:band28FM}a and \ref{FIG:band28FM}b, where
we chose to plot the mean-field band structure for the ground-state solution at $x=2.8$, $n_c\approx0.5$ and $n_c\approx0.67$. The high value of the coupling
has been chosen to give a better visualization of the features of the FM-I phase; anyway all the ground states in this phase share the same characteristics.
Of course at mean field level, considering our assumptions (\ref{eq:conditionsSYMMETRY}),  the entanglement of the two many-body wave-functions is lost,
so the partially filled $g$-band represents (effectively) the band formed by the majority-electrons that are not bound into the Kondo-singlets.

\begin{figure}[t]
\includegraphics[width=1\columnwidth]{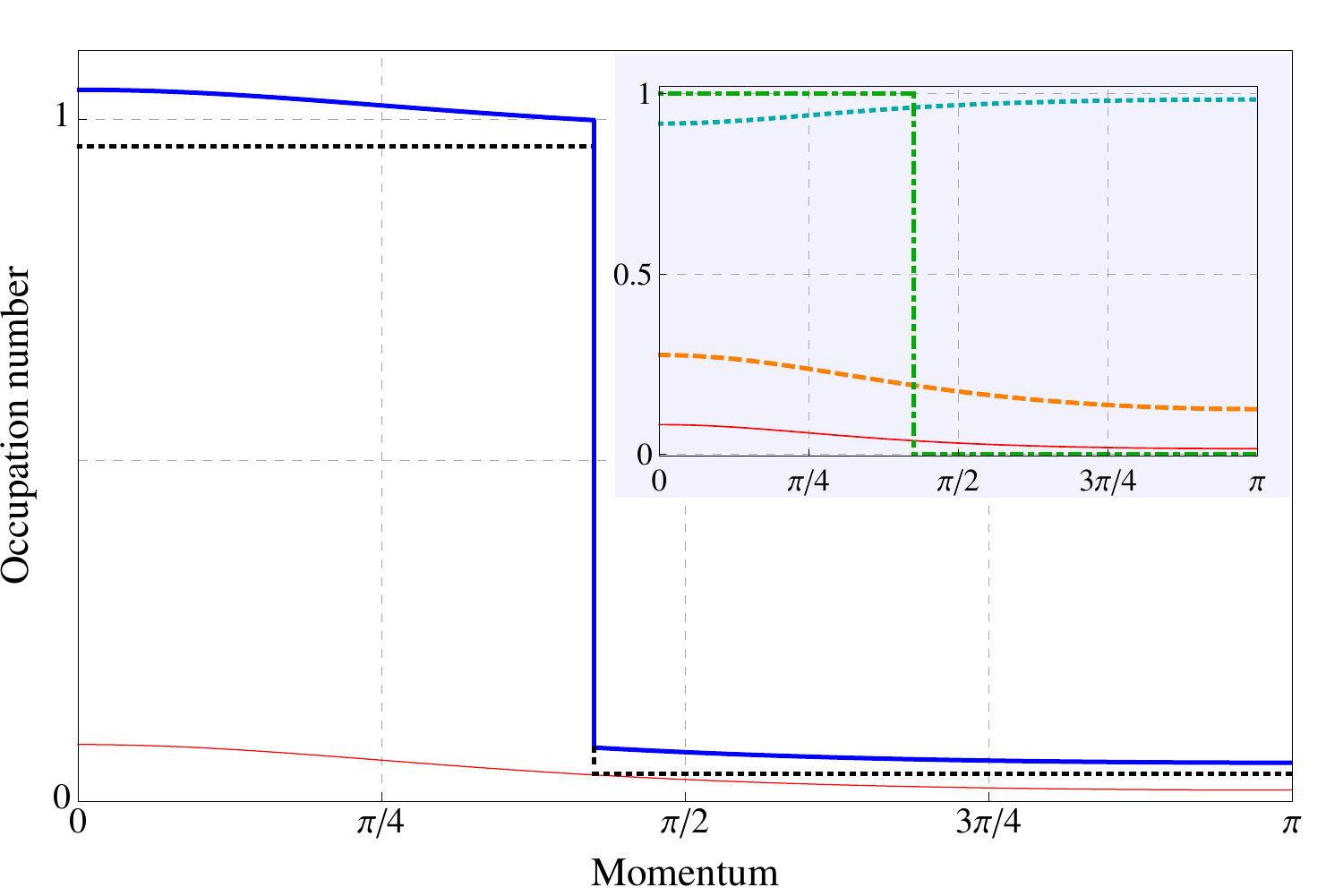}
\caption{Expectation values on the FM-I $cgf$ mean-field ground-state at $x=2.8$ and $n_c\approx0.5$, corresponding to the band structure in Fig.~\ref{FIG:band28FM}a.
MAIN: Momentum distribution on the Brillouin zone of the operators $n_{c,\ua}(k)=\langle c^\dagger_{c,\ua(k)}c_{c,\ua(k)}\rangle$ {\it thin red},
$n_{c,\da}(k)=\langle c^\dagger_{c,\da}(k)c_{c,\da}(k)\rangle$ {\it dotted black} and $n_{c,tot}(k)=n_{c\da}(k)+n_{c,\ua}(k)$ {\it thick blue}.\label{FIG:c(k)c(k)}
INSET:
For the same state, the momentum distribution on the Brillouin zone of the operators $\langle g^\dagger(k)g(k)\rangle$ {\it green dotdashed},
$\langle-i(f^\dagger(k)c(k)-c^\dagger(k)f(k))/2\rangle$ {\it dashed orange}, $\langle f^\dagger(k)f(k)\rangle$ {\it dotted cyan} and $\langle\hat n_{c_\ua}(k)\rangle$ {\it thin red}.
Both figures are symmetric for $k\rightarrow-k$.\label{FIG:meanfieldFM28}
}
\end{figure}

The previous description gives a qualitative rationale for the unit value of the commensurability.
The idea is that there must occur a fine-tuning between the density of the minority-electrons and of the minority-spins.
Indeed, the creation of minority-spins is energetically expensive: for any majority-spin turned into a minority-spin
there is a loss in the kinetic energy
of the $g$-fermions, because of the reduction of the mean-field $\mathcal F$.
Therefore the density of the minority-spins will be as small as possible, i.e. there will be an equal number of minority-spins and minority-electrons.

It is now clear that in our description, while the $c$-fermions represents the minority-electrons, the $f$-fermions represent the majority-spins; thus
the vacancies in the completely filled $f$-band must represent the minority-spins and hybridize with the $c$-fermions.
Imposing the same value for the densities of the minority-electrons and minority spins, remembering that the average number
of minority-electrons is $\mathcal C$ and that the average number of minority-spins is $1-\mathcal F$, it is straightforward to obtain:
$$
\mathcal C=1-\mathcal F\quad \Rightarrow \quad \mathcal C+\mathcal F=1.
$$
This is exactly the justification provided in Ref.~\onlinecite{Peters:2012fc} for the unitary value of the commensurability parameter, given in
terms of $cgf$-fermions. As is evident our $cgf$-formalism fits perfectly the physics of the SSKI and therefore the FM-I phase.

\begin{figure}[t]
\includegraphics[width=1\columnwidth]{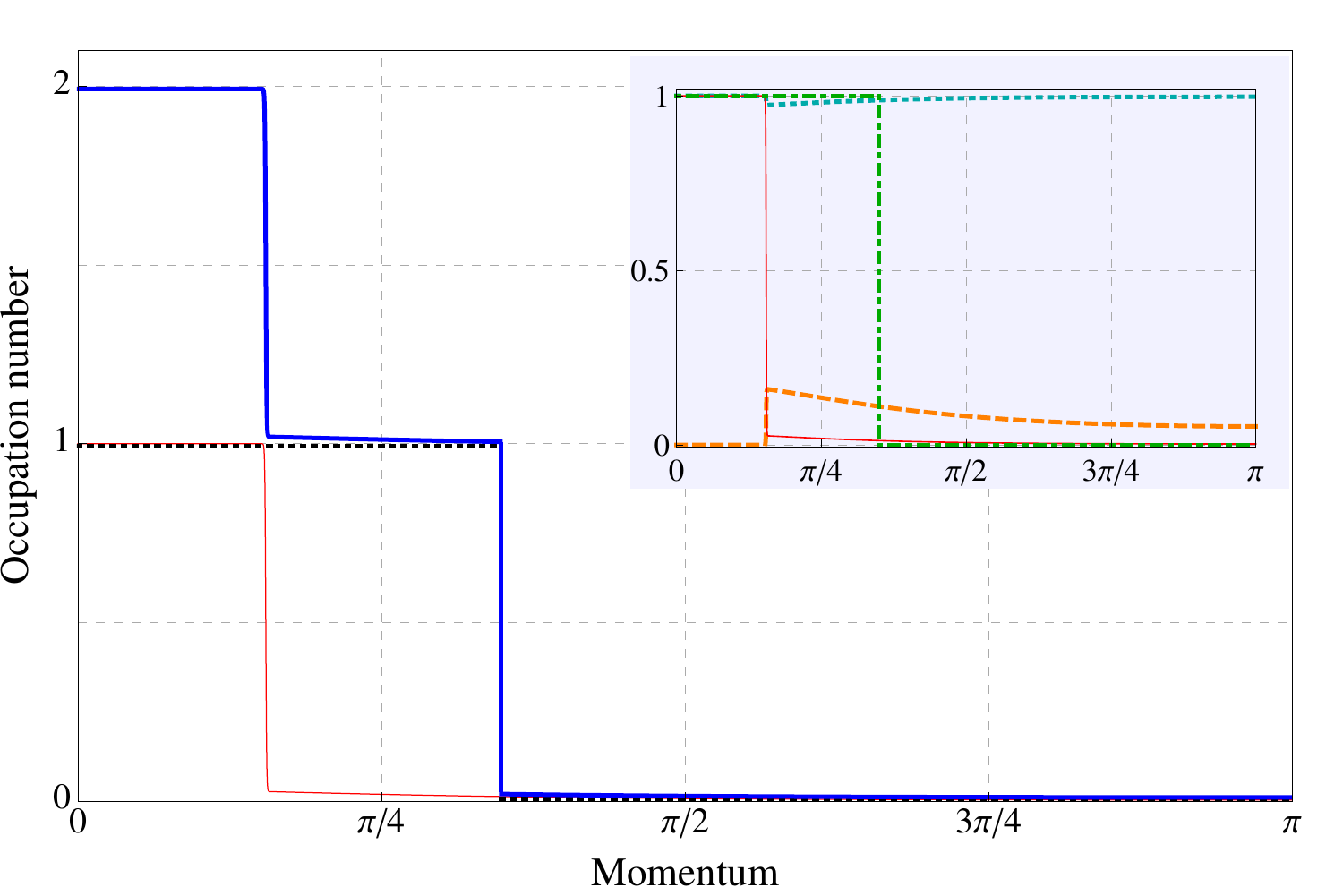}
\caption{Expectation values on the FM-II $cgf$ mean-field ground-state at $x=1.4$ and $n_c\approx0.5$, corresponding to the band structure in Fig.~\ref{FIG:band28FM}c.;
MAIN: Momentum distribution on the Brillouin zone of the operators $n_{c,\ua}(k)=\langle c^\dagger_{c,\ua(k)}c_{c,\ua(k)}\rangle$ {\it thin red},
$n_{c,\da}(k)=\langle c^\dagger_{c,\da}(k)c_{c,\da}(k)\rangle$ {\it dotted black} and $n_{c,tot}(k)=n_{c\da}(k)+n_{c,\ua}(k)$ {\it thick blue}.\label{FIG:densitiesFMII14}
INSET:
For the same state, the momentum distribution on the Brillouin zone of the operators $\langle g^\dagger(k)g(k)\rangle$ {\it green dotdashed},
$\langle-i(f^\dagger(k)c(k)-c^\dagger(k)f(k))/2\rangle$ {\it dashed orange}, $\langle f^\dagger(k)f(k)\rangle$ {\it dotted cyan} and $\langle\hat n_{c_\ua}(k)\rangle$ {\it thin red}.
Both figures are symmetric for $k\rightarrow-k$.\label{FIG:fieldsFMIIx14}
}
\end{figure}

We plot an enlightening outcome of our $cgf$-map mean-field analysis in Fig.~\ref{FIG:c(k)c(k)}. The curves represent
the momentum distribution of the conduction electron density operators.
As can be seen the Fermi surface of the $c_{c,\ua}$ minority-electrons
is completely destroyed by the hybridization and as a matter of fact the minority-electrons
are not expected to show any Fermi-liquid behavior, because they exist {\it only as components} of the coherent Kondo singlets and {\it not as free particles}.
The only jump is visible in the majority-electron distribution (and consequently in the total distribution also).
The fact that a part of the majority electrons participate in the formation of the Kondo singlets is made evident by the fact that
their occupation number is not equal to one inside the Fermi volume.
The position of the Fermi momentum is compatible with the picture presented, where almost all the electrons are stored in the $c_{c,\da}$ (majority) band and only
those that are not bounded into singlets contribute to the Fermi volume. This explains what is the nature of our mean field solution: it separates
the part of the majority-electron wave function and the spin-singlets one, i.e., it stores the effective fraction of the majority electrons that can be thought
as free in the $g$ sector. The more they are, the less the Kondo singlet wave function is entangled with the majority electrons one.
We want to point out how our description is not only able to move the Fermi momentum correctly, but also to renormalize the
jump at the Fermi surface. This happens because the creation of the Kondo singlets spreads part of the electron-quasiparticle weight over the
entire Brillouin zone.
For future convenience
we also plot the momentum distribution of some $cgf$-operators in
in Fig.~\ref{FIG:meanfieldFM28} (inset). As can be seen the hybridization takes place at every momenta,
indicating that all quantum states of the minority electrons are involved into generation of the Kondo singlets.


In the FM-I phase the double exchange effect clearly dominates and the Kondo effect enters as a way to optimize the energy, permitting the
``annihilation'' of the unwanted minority-electrons, but without creating any global Kondo singlet state.
In spite of that, increasing more and more the filling, the status quo does not survives up
to half-filling. When a critical value $n^F_{crit}(J)$ is reached, the physical properties of the system change completely, passing
from the FM-I to the FM-II phase. As mentioned earlier this happens when also the $c$-like band crosses the Fermi level at zero energy and starts to get filled. The filling
of the $c$-like states is not the only thing that changes in this process. In fact the states that get filled do not contribute anymore to the hybridization field $\mathcal{I}$, as can be
seen in Fig.~\ref{FIG:fieldsFMIIx14} (inset), that is an example for $x=1.4$ and $n_c\approx0.5$, corresponding to the band structure of Fig.~\ref{FIG:band28FM}c.
The hybridization field $I(k)=\langle-i(f^\dagger(k)c(k)-c^\dagger(k)f(k))/2\rangle$ is zero where the $C(k)=\langle c^\dd(k)c(k)\rangle$ is one,
indicating that the filled states are well defined $c_{c,\ua}$-electron states, characterized by no charge fluctuations, in contrast
with the FM-I ground state where no $c_{c,\ua}$ state was fully occupied, because all were involved in the Kondo singlets formation.

In terms of the $c_{c,\ua}$ (minority) and $c_{c,\da}$ (majority) electrons what happens is that for $n_c>n^F_{crit}(J)$ the system is not anymore able
to keep such an high unbalance between the two electrons species and tries to equilibrate the two populations. Some minority electrons
escape the process of bonding into the Kondo singlets and so are allowed to hop freely from site to site. This is manifested by the creation of a Fermi surface also for the
minority $c_{c,\ua}$ electrons, that remarkably appears in the FM-II phase, as shown in Fig.~\ref{FIG:densitiesFMII14} for illustration purposes.

The physics of the FM-II phase is much less exotic than the one described by the FM-I phase. In fact, since the hybridization is less pronounced and
both the fermion species have a Fermi surface, it is easy to relate this phase with that of an electron liquid polarized by the
magnetic field generated by the impurity spins, i.e. what was previously called RKKY-ferromagnet and that corresponds to the mean field
ferromagnetic state considered in the standard literature\cite{Fazekas:1991kc}. For example in Fig.~\ref{FIG:band28FM}c and \ref{FIG:band28FM}d
it is very evident how the $c$-like and $g$ bands roughly represent
the two $c_{c,\ua}$ and $c_{c,\da}$ bands. In this scenario the hybridization term, remnant of the SSKI formation of the FM-I phase, optimizes the
ground state configuration incorporating the Kondo effect into it. This optimization is very efficient and in fact the FM-II phase
gains a lot of energy when the coupling is increased, beating also the mean field spiral ordered ground states\cite{Fazekas:1991kc}.
The FM-II phase is related to the RKKY ferromagnetic one, in the sense that for low filling and low coupling these two kind of ground states
look very much the same, becoming more and more different increasing the relevancy of the Kondo effect.

It is important to point out how the FM-II phase survives from $n^F_{crit}(x)$ up to half-filling, for each $x\lesssim2$.
Of course this does not mean that it represents the true phase of the system, and in fact
it is known that the region of the phase diagram occupied by the FM-II phase is mostly paramagnetic and not ferromagnetic. It is therefore
very probable that comparing the FM-II ground states with mean field solutions that incorporate, beside the Kondo effect, also a non-translational invariant
order (i.e. the analogous of the FM-II states but with spiral order for the spins), then the FM-II states would not be favorable anymore.
Anyway the numerical simulations\cite{McCulloch:2002ss} clearly indicate the existence of a ferromagnetic phase, a sort of ferromagnetic tongue, {\it inside}
the paramagnetic dome. To the best of our knowledge there exist no theoretical explanation for this tongue, whose ferromagnetism
has never been fully understood. The FM-II states, able to dominate the spiral spin ordered trial states, provide a possible justification
for the ferromagnetic tongue, and could be regarded as prototype for this kind of Kondo stabilized ferromagnetic order.

\subsubsection{Instability of the FM-I phase at intermediate couplings}
\begin{figure}[t]
\includegraphics[width=1\columnwidth]{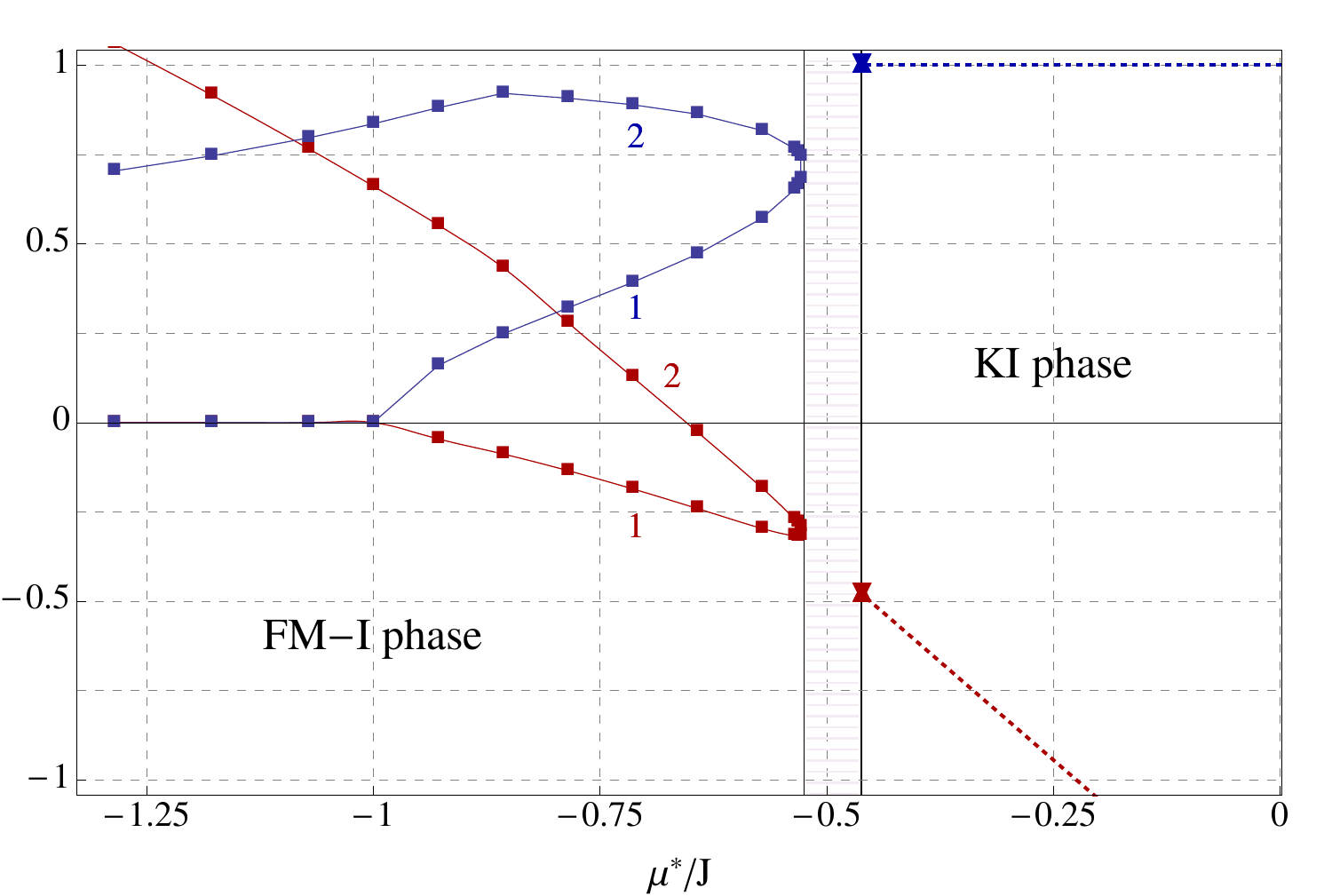}
\caption{Free energy, {\it red}, and (on-site) electron density, {\it blue}, versus $\mu^*/J$, for $x=2.8$. On the y-axis the free energy is given in
units of $4t/\pi$ and the value $0$ and $-1$ correspond respectively to the empty and the half-filled
non-interacting model; the density can vary from zero (no electrons) to one (half-filling).
The labels 1 and 2 mark the low- (FM-I ground-state) and high-energy mean field solutions. The label KI indicates the half-filled Kondo insulating solution.
The shaded area, delimitated by the two vertical black lines, indicates the values of the chemical potential where the bifurcation takes place.
The dots of the branches 1 and 2 indicate the solutions that we found numerically, while the lines are an interpolation of the results. In the case of the KI
solutions instead, we could obtain an arbitrary large amount of points, therefore the dashed KI line is drawn with machine precision.
The error bars are not present, because they are smaller than the marker points.
\label{FIG:freeenergy28}
}
\end{figure}

At $x\approx2$ it appears an instability in the phase diagram. In fact the two branches (low and high energy, mentioned at the beginning of Sec.~\ref{sec:lowcoupling})
of mean field solutions collide for a critical chemical potential $\mu_{12}(x)$,
corresponding to a critical density $n^{pol}_{crit}(x)$. This collision implies the disappearance of both the solutions, that merge into a new one: the half filled KI.
This solution is the the only translational invariant mean field solution for $\mu^{pol}_{crit}<\mu^*<0$ and it
appears at $\mu^{pol}_{crit}>\mu_{12}(x)$ leaving an interval between $\mu_{12}(x)<\mu^*<\mu^{pol}_{crit}$ (shadowed region in Fis. \ref{FIG:freeenergy28}
and \ref{FIG:commensurab,spin mag,electronmag28}) where no solution could be found.
In turn, this also means that we found no solutions between $n^{pol}_{crit}<n_c<1$. Although
strange this result has a quite reasonable explanation, consistent with the physics of the one-dimensional Kondo lattice.

To understand this process
of collision of two solutions, one has to keep in mind the nature of the
mean field solutions, which are fix points of our Newton-Raphson method, or also extremal points
of the mean field energy functional (see Appendix \ref{app:A}). The only way for fix points (or two extrema) to disappear is via a bifurcation process.

Initially at $x\approx 2$, we found that $n^{pol}_{crit}\approx 1$; changing
the coupling this critical density moves quickly to much smaller values, so that at $x\approx2.2$ one has already $n^{pol}_{crit}\approx 0.8$. In doing this, the critical density
becomes smaller than the critical density that separates the FM-I and FM-II phases, making the latter disappear from the phase diagram. 
So at intermediate couplings $x\gtrsim 2.1$, the FM-II phase does not exist anymore and we found only two phases: the FM-I phase between
$0<n_c<n^{pol}_{crit}$ and the KI phase at $n_c=1$.
We would like to point out how the value of the coupling where the FM-II phase disappears is in approximate agreement with the upper boundary of the
ferromagnetic tongue phase mentioned previously, suggesting again the connection between the two phases.

The absence of a solution between $n^{pol}_{crit}<n_c<1$ is an annoying feature, but it hides a possible physical explanation. The picture
becomes more clear analyzing the behavior of the mean field solutions at varying chemical potentials, rather than varying density.
In Fig.~\ref{FIG:freeenergy28} we plot the dependence of the free energy and electron density on the chemical potential, for our mean-field solutions,
choosing $x=2.8$ for illustrative purposes (the same structure holds for all the intermediate couplings).
In Fig.~\ref{FIG:commensurab,spin mag,electronmag28} we plot also the behavior of other physical quantities,
such as the value of the commensurability parameter, impurity spin polarization
and electron polarization.

\begin{figure}[t]
\includegraphics[width=1\columnwidth]{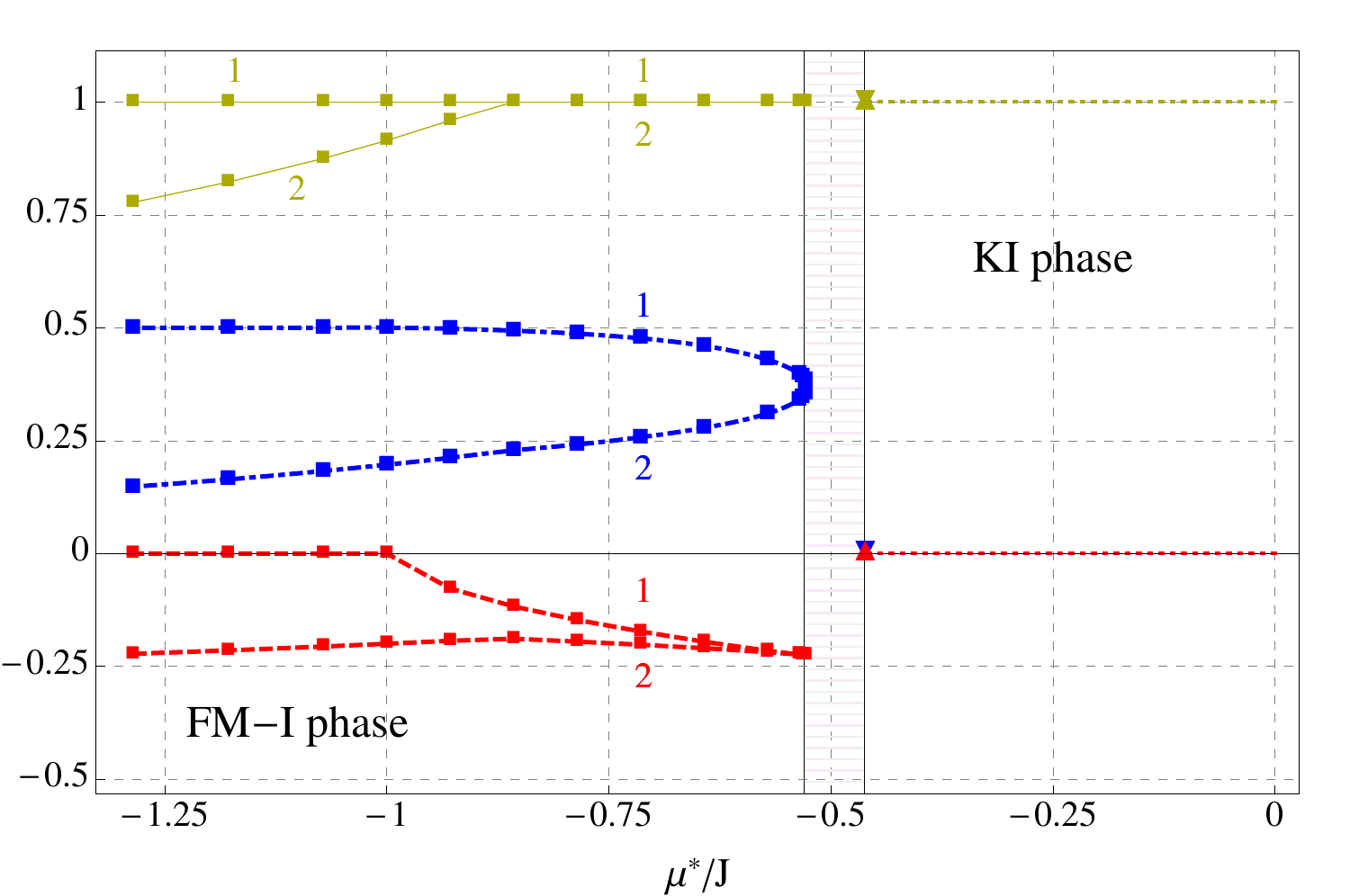}
\caption{Commensurability, {\it top-yellow}, impurity spin magnetization $\langle S_f^z \rangle$, {\it middle-blue},
and electron spin magnetization $\langle c^\dagger_{c,\ua} c_{c,\ua}-c^\dagger_{c,\da} c_{c,\da}\rangle/2$, {\it bottom-red}, versus $\mu^*/J$, for $x=2.8$.
The same conventions as in Fig.~\ref{FIG:freeenergy28} are used. It is evident how the commensurability
is always equal on the ground-state of the FM-I phase (i.e. the branch labeled by the number 1).
\label{FIG:commensurab,spin mag,electronmag28}
}
\end{figure}

On the left can be seen the two branches of mean field solutions. As evident the two families become degenerate at $\mu_{12}=\mu^*/J\approx -0.53$,
and converge to the same point in the parameter space, as can be understood examining the electron density curve (we reserve
the symbol $\mu^{pol}_{crit}$ for the value of the chemical potential at which we are able to resolve the new KI phase).
Increasing the chemical potential we were not able to resolve any solution until the appearance of the KI phase at $\mu^{pol}_{crit}$.
This is due to the fact that the mean field energy functional becomes
almost flat, making impossible the identification of maxima, minima and flexes for values of $\mu^*$ between $\mu_{12}$ and $\mu^{pol}_{crit}$.
The flat shape of the mean field energy functional is the result of the collision between the two branches (fix points)
and physically it has a quite natural interpretation, visible in Fig.~\ref{FIG:freeenergy28}. Indeed, following the ground state branch $1$, i.e., the FM-I phase,
it is evident that the function $n_c(\mu^*)$ is going towards a vertical flex at $\mu_{12}$, which also means that the derivate $dn_c(\mu^*)/d\mu^*$ diverges at $\mu_{12}$.
Since $dn_c(\mu^*)/d\mu^*$ is proportional to the compressibility\cite{NozieresPines} of the quantum liquid its divergence signals an instability and a phase transition
due to a process of phase separation.
At $\mu_{12}$, corresponding to the density $n_{crit}^{pol}$ the energy necessary to add an electron becomes zero. In terms of our algorithm the
divergence of the compressibility is manifested by the divergence of some elements in the Jacobian matrix, and so in the impossibility
to resolve the fixed points in the shadowed region in Fig.~\ref{FIG:freeenergy28} and \ref{FIG:commensurab,spin mag,electronmag28}.

With this in mind we can try to explain the physics behind this behavior.
As mentioned in the previous section, for small values of density (i.e for low chemical potential) the electrons are able to delocalize on the entire lattice, creating a coherent
magnetization on the entire system and generating the state FM-I that survives for higher and higher densities,
stabilized by the creation of the SSKI that can be though of as a liquid of Kondo singlets.
However, once that the critical density $n^{pol}_{crit}$ is reached at $\mu_{12}$,
the FM-I is not anymore able to host new electrons and small ``bubbles'' of the half-filled KI phase appear
in the system, separating {\it islands} of FM-I phases that become less and less extended increasing the total electron density.
In these islands the ferromagnetic order is still realized by the conduction
electrons, via double exchange. With respect to the SSKI picture discussed previously,
one understands that a qualitative two-liquid picture can be elaborated to take into account the physics of the system:
in the FM-I phase below the critical density $n^{pol}_{crit}$ the two liquids (the majority-electrons liquid and the Kondo-singlets liquid)
are homogeneously mixed on the entire lattice and their wave functions entangled.
When the critical $n^{pol}_{crit}$ is reached, it is not energetically favorable to keep this homogeneous configuration and the two fluids separate.
This phase separation is marked by the divergence of the compressibility.

This picture of the phase separated region resembles the
description provided by bosonization\cite{Gulacsi-:2004nx,Honner:1997qo}, where the islands
of coordinated spins are identified as polarons and the phase coexistence region is the polaronic liquid.
In this region of the phase space the correct degrees of freedom are\cite{Gulacsi-:2004nx} the islands of FM-I phase,
or more properly the electrons dressed by the ferromagnetic polarized cloud of impurity spins.
Our {\it mean-field analysis is not able to describe} the dynamics of these polarons, but permits to
predict their existence and locate, at intermediate couplings, their liquid phase in the correct region\cite{Peters:2012fc,McCulloch:2002ss}
of the phase diagram, although not perfectly.

Clearly the impossibility to follow the mean field ground state into the polaronic liquid region is a feature of our
mean field decomposition scheme (\ref{eq:conditionsSYMMETRY}). In principle, allowing for more hybridization channels, also
an analysis of the polaronic liquid would be possible, but this would spoil the advantages of the $cgf$-map, making the solution very involved.
It is our opinion that, if the subject of the study are the properties of the polaronic liquid, i.e., of the heavy fermion phase of the one-dimensional Kondo
lattice\cite{Gulacsi-:2004nx}, then it would be more appropriate to modify the mapping. This unsuitability of the $cgf$-map, as we have defined it,
in the mean field description of the polaronic phase is consistent with the fact that we optimized the mapping for the analysis of ferromagnetic (or
in general translational invariantly ordered) states.

\subsubsection{The scenario at high couplings}
Increasing even more the coupling, reaching $x\gtrsim3$, the scenario does not change, except for the fact that $n_{crit}^{pol}$
moves to lower and lower values. Moreover some other unphysical mean field solutions appear close to half-filling. We believe that
these are symptoms of the fact that for such an high value of the coupling the ground state structure changed too much, with respect to the
original one. The FM-I phase stabilized by the SSKI mechanism (as we explained it previously, in the mean field picture) does not give
anymore a good approximation of the ground state configuration. This is obviously due to the enhanced importance of the Kondo effect,
that causes a stronger and stronger entanglement of the majority-electron wave-function with the Kondo-singlets one. Eventually
there is not anymore space to think of a part of the majority $c_{c,\da}$-electrons as free, i.e. as a $g$-fermion sector
completely decoupled from the $c$-$f$ one, and our mean-field decomposition scheme breaks down.

Anyway this arguments suggests the possibility for the existence of two qualitatively different ground states describing the
ferromagnetic phase of the one-dimensional Kondo lattice at low and high coupling.

\subsection{Half-filled solution: the KI state and the RKKY effect}\label{subsec:MS}

It is well documented\cite{Tsunetsugu:1997vn} that half-filling is a very special point for the Kondo lattice model. The configuration
of the ground-state is very different from the ones that are infinitesimally close to it, in particular for what concerns the
magnetic properties of the system.

At half-filling the system forms a spin liquid with total spin $S=0$, characterized
by a gap in both the spin and charge sector. The gaps exists for every value of the coupling, and no critical $x$ that signals
a phase transition has ever been found, although it is strongly believed that the mechanisms responsible for the existence
of the gap are different in the two limits.

At small coupling the RKKY effect causes a local antiferromagnetic\cite{Tsunetsugu:1997vn,Tian:1994pi} order in the impurity-spins. This order is only local
and quantum fluctuations destroys it at larger scales, implying the opening of the spin gap\cite{Le-Hur:1998ly,Tsvelik:1994hs}. However the electrons moving on the lattice
feel the nearest neighbor antiferromnagnetic order, experiencing coherent Bragg (back-)scattering and a gap opens also in the charge sector\cite{Le-Hur:1998ly}.
At high coupling the nature of the gap is instead caused by the development of the local Kondo singlets. This gap is much similar to the BCS gap
of superconductors\cite{Eder:1997kl}: it opens because a local singlet has to be
broken to move the local charge or flip a local spin, costing an energy of $3J/4$.
Since no phase transitions between the two regimes exists, a cross-over\cite{Tsunetsugu:1997vn} must take place around some value $x$.

It is quite clear that a mean-field approach will not be able to capture correctly the subtle physics of the spin liquid phase. As a matter of fact we already tried
to tackle the problem in more interesting and appropriate way\cite{ChristmaMajorana}, keeping the spin-rotational symmetry and studying
the sector of non-magnetic ground-states.
Anyway a discussion
of the mean-field results will not be completely meaningless, because some interesting features are correctly captured by the mean-field solutions. Moreover
it will be a good occasion to discuss the appearance of the RKKY effect in the context of the $cgf$-map.

In our mean-field phase diagram, the particularity of the half-filled point is the existence of the KI state. Such mean-field solution exists only at half-filling, like
a singular point. It is characterized by perfect\footnote{Within our numerical precision, that in this case is fixed arbitrarily in $10^{-9}$.}
paramagnetism $\mathcal F=1/2$ and perfect balance between the up-down population of the conduction electrons $\mathcal C=1/2$.
The value of the mean-field $\mathcal I$, that measures the average hybridization between the species $c$ and $f$, is coupling dependent and
goes from zero at $x\rightarrow 0$ to $1/2$ at $x\rightarrow +\infty$.

\begin{figure}[t]
\includegraphics[width=1\columnwidth]{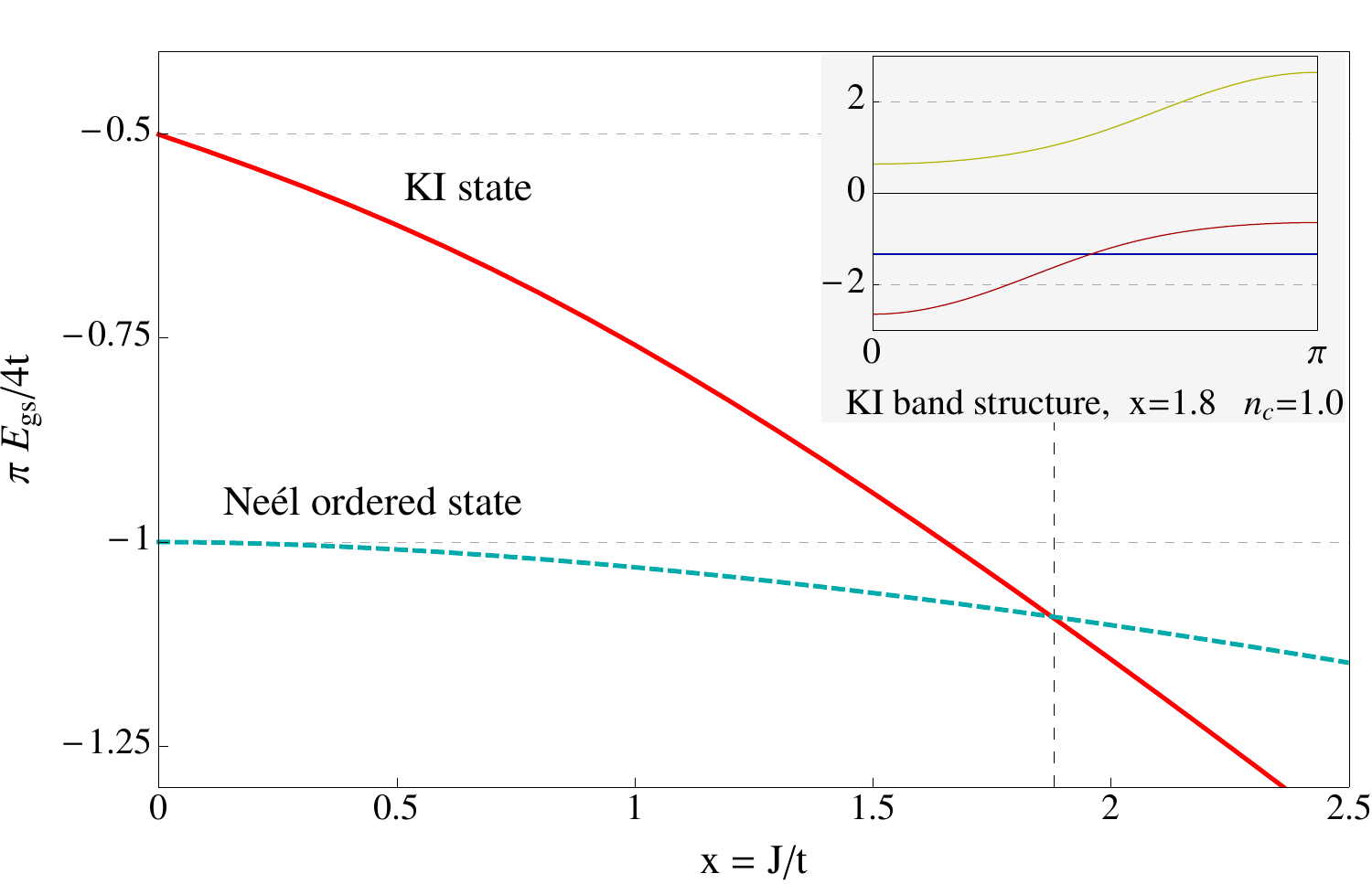}
\caption{MAIN: The evolution of the mean field energy of the state KI, compared with the antiferromagnetic spin ordered ground state of Ref.~\onlinecite{Fazekas:1991kc}.\label{FIG:fazekasVSus}
INSET: the cgf band structure of the KI ground state for the value $x=1.8$; on the y-axes the energy is given in units of $t$. The same conventions of Fig.~\ref{FIG:band28FM}
have been kept; the bands are symmetric for $k\rightarrow -k$.}
\end{figure}

Of primary importance is the fact that $\mathcal G=1$ for every value of the coupling.
The fact that the $g$-band is completely filled means that there is always one $g$-fermion
per site and therefore the available states for the description of the mean-field KI state are only (see Tab.~\ref{tab:cgfmap}):
$$
|\Downarrow\rangle,\quad |\ua \Downarrow\rangle,\quad |\da \Uparrow\rangle, \quad |\ua\da\Uparrow\rangle.
$$
The presence of the states with zero and two electrons seems annoying, but it is necessary in order to keep in the ground state wave function
also an uncertainty in the local conduction electron density. Indeed only at $J\rightarrow +\infty$, when all the conduction electrons bound into locally inert Kondo singlets,
the local conduction electron density is exactly equal to one.
As can be seen in Fig.~\ref{FIG:fazekasVSus} (inset), the mean-field KI solution is characterized by two bands (together
with the flat filled $g$-band) that comes from the originally flat $f$-band, hybridized with the originally cosine-shaped $c$-band.
The hybridization gives to the two bands the avoided-crossing structure typical of Kondo insulators. This band structure does not
change if the chemical potential is modified, as long as it remains inside the gap. When the critical chemical potential $\mu^{pol}_{crit}$
is reached, i.e., when the chemical potential level intercepts one of the bands, the KI solution collapses and a new couple of solutions (the two FM-I solutions
that crush at $\mu_{12}$) become the only mean-field solutions.

Given that $\mathcal G=1$, the two bands of the half-filled KI state are found diagonalizing the $H^{MFTI}_{cf}$. Defining two creation operators as:
\be
s^\dagger(k)=\sin(\theta_k)c^\dagger(k)+i\cos(\theta_k)f^\dagger(k),\nonumber\\
t^\dagger(k)=\sin(\theta_k)c^\dagger(k)-i\cos(\theta_k)f^\dagger(k),\nonumber
\ee
then the KI solution is given by the ground state
\be
&|KI(x)&\rangle=\prod_{k=-\pi}^\pi s^\dagger (k)g^\dagger(k) |0_{cgf}\rangle\\
&=&\prod_{k=-\pi}^\pi \left(\sin(\theta_k)c^\dagger(k)+i\cos(\theta_k)f^\dagger(k)\right) g^\dagger (k)|0_{cgf}\rangle,\nonumber
\ee
where the $x$ dependence enters into the the functions $\theta_k$.

For $x\rightarrow +\infty$ the KI state approaches the correct asymptotic ground-state with $\mathcal I=1/2$, i.e.
$$
|KI(x\rightarrow+\infty)\rangle=\prod_{k=-\pi}^\pi \frac{1}{\sqrt{2}}\left(c^\dagger(k)+if^\dagger(k)\right) g^\dagger(k) |0_{cgf}\rangle.
$$
The fact that $|KI\rangle$ is the correct mean-field ground-state for $x\rightarrow +\infty$ can be understood also without any numerical
analysis looking at $H^{MFTI}_{cgf}$ putting $t=0$ and sending $J$ to infinity.
Approaching the correct infinite-coupling ground-state it is not surprising that also the correct asymptotic energy density dependance of $-3x/4$ is recovered.

It is important to note that in this limit the ground state is correctly given by a linear combination of Kondo singlets: one for each site.
In fact on each site one has the realization of the state $s^\dagger g^\dagger |0\rangle$ that means $(c^\dagger +if^\dagger) g^\dagger |0\rangle$
that by the $cgf$-map Tab.~\ref{tab:cgfmap} is $|\ua\Downarrow\rangle-|\da\Uparrow\rangle$. In the case $x<+\infty$ the $k$-dependence of $\theta_k$
spoils the singlets with components coming form the states $|\Downarrow\rangle$ and $|\ua\da\Uparrow\rangle$, necessary to take into account the hopping of the electrons; while
the fact that $\theta_k\neq\pi/4$ implies also the contribution of the triplet component with spin $S_{tot}^z=0$. These properties are
in agreement with the known high coupling solutions\cite{Eder:1997kl,Eder:1998fu}.

\begin{figure}[t]
\includegraphics[width=1\columnwidth]{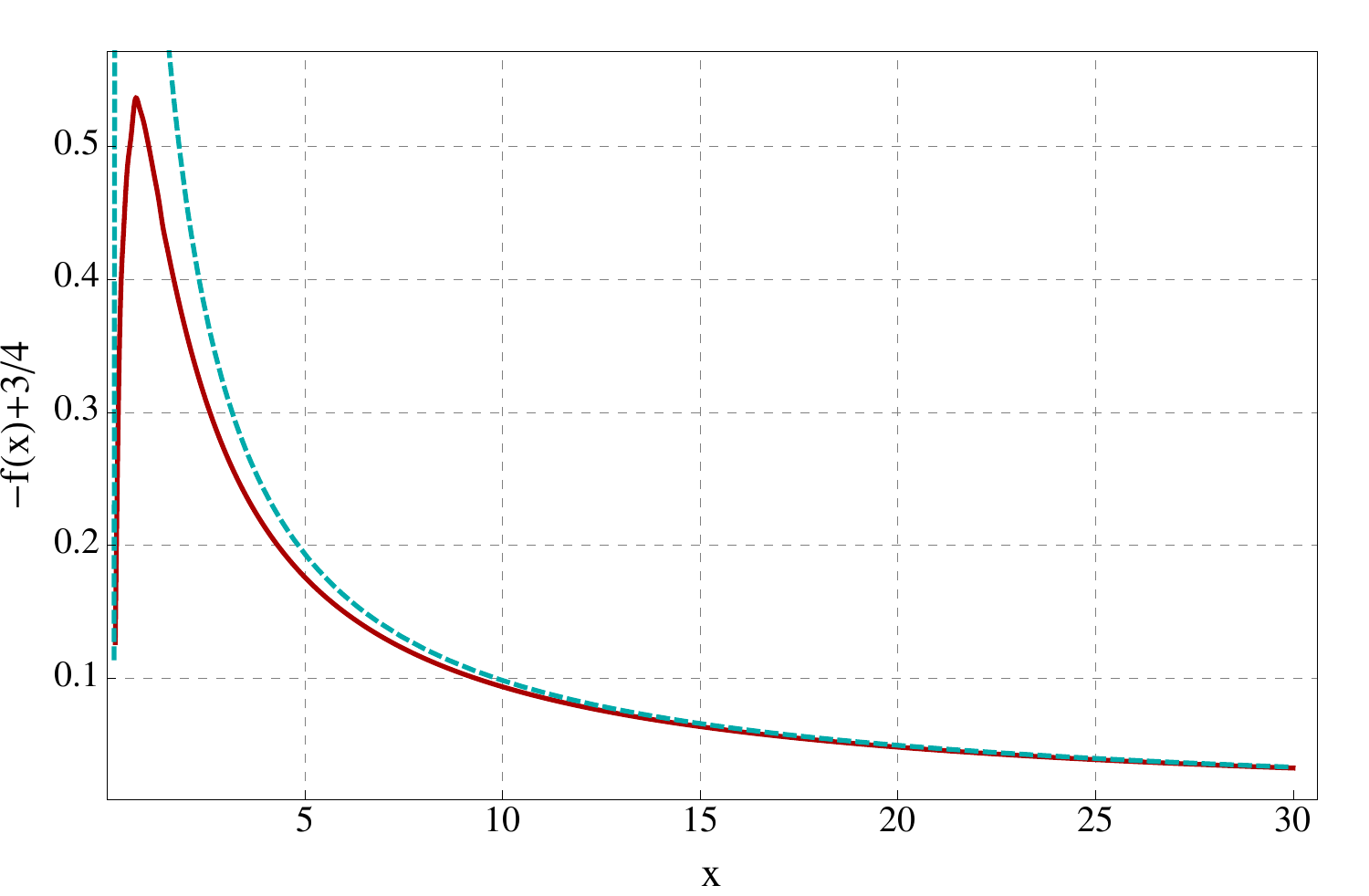}
\caption{The different curves are the numerically determined $f(x)=\mu^{pol}_{crit}(x)/J$, {\it continuos red}, and $f(x)=\Delta_{qp}(x)/J=1/6x^2-1/x+3/4$, {\it dashed blue},
that is the perturbative $t/J=1/x$ expansion for the quasiparticle gap.
\label{FIG:mucrit}
}
\end{figure}

The mean-field gap between the two bands, that corresponds at infinite coupling to the gap between the singlet and triplet states at $S_z^{tot}=0$,
is equal to $3x/2$. Unfortunately this mean-field gap does not agree with the correct spin-gap of the Kondo Insulator solution, that should be equal to $x$
in the high coupling limit. However this is not so surprising, because one cannot expect to predict properties of the excited states using a
trivial (time-independend) mean-field theory. By construction, the critical chemical potential $\mu^{pol}_{crit}$ corresponds to
the energy necessary to add or remove one particle from the system. This energy has been already defined in the KLM as the quasiparticle gap;
we compare the value of $\mu^{pol}_{crit}$ and of the quasiparticle gap, using the known\cite{Tsunetsugu:1997vn} high coupling perturbative expansion
to compute it. As evident in Fig.~\ref{FIG:mucrit} the asymptotic behavior at high coupling is the same. 
Anyway around $x\approx 10$
a qualitative change in the behavior of the gap  is expected\cite{Tsunetsugu:1997vn}, due to the non negligible effect of the RKKY interaction, therefore
both the curves are not relevant below that value of the coupling.
The inadequacy of the state $|KI(x)\rangle$ at small coupling is made evident by the fact that for $x\lesssim2$
it is neither the energetically most favorable solution among the translationally invariant ones, because also the FM-II phase exists at half-filling.
An analysis of (\ref{eq:MFHamiltonianTI}) immediately reveals the problem: at small $x$ clearly
the system prefers the strongly ferromagnetic order to the paramagnetic one because the kinetic energy contribution of the $g$-fermions
gets maximized (recall that in the FM-II phase the $g$-fermions can be interpreted as the $c_{c,\da}$-electrons).
So as long as one considers only translational invariant solutions, the ferromagnetic order at small coupling is not avoidable.

However, if one instead looks at (\ref{eq:MFHamiltonian}) it becomes clear that there exist a way to recover the
kinetic energy of the $g$-fermions, without implying ferromagnetic ordering of the impurity spins. In fact it
assuming (perfect) {\it antiferromagnetic} order, one obtains for $H_g$:
\be\label{eq:HantiG}
H^{AF}_g&=&t\left(1-2\mathcal F \right)\sum_{n} \left(g^\dd \tilde g^\dd + \tilde gg\right)+\frac{J}{4}(-1-4  \mathcal I)\sum_{n} g^\dd g,\nonumber
\ee
where $\mathcal F$ is the mean-field on the first site of the double unit-cell and we assumed $\mathcal I=const$ for sake of simplicity.
This hamiltonian can be solved in many ways, for example doubling the unit cell, mapping $\tilde g\rightarrow a^\dagger$ and making use of Nambu spinors.
It is clear that the contribution from the kinetic energy is equally well obtained and so this state will approach the energy value of the half-filled zero-coupling
solution as well as the ferromagnetic one.
This can also be understood using (\ref{eq:g-cgfmap}): assuming saturated impurity spin ferromagnetism then the $g$-fermion operators are
the $c_{c,\da}$ operators and $g^\dagger=c^\dagger_{c,\da}$;
assuming instead saturated spin antiferromagnetism we have $g^\dagger=c_{c,\da}^\dagger$ on the sites with spin up and $g^\dagger=c_{c,\da}$
on the sites with spin down. Therefore the hopping term $g^\dagger \tilde g^\dagger$ in the antiferromagnetic case is exactly $c_{c,\da}^\dagger \tilde c_{c,\da}$.

To study the competition of the antiferromagnetic and the ferromagnetic ground-states we should solve the mean-field Hamiltonian (\ref{eq:MFHamiltonianTI})
imposing translational invariance on the doubled unit cell.
This is not the analysis that we carried on.

The imposition of perfect antiferromagnetism, i.e.
$\mathcal F_{n+1}=1-\mathcal F_n$ with $\mathcal F_n=0$,
implies that no hybridization is possible
between the $c$ and the $f$ fermions (otherwise the value of the $\mathcal F$ field would be spoiled), hence $\mathcal I=0$.
This also means that all the interesting features of our model are neglected, as the Kondo effect,
and only the RKKY effect is kept into consideration.
The solution for the energy of the antiferromagnetic ground-state at half-filling is the well known\cite{Fazekas:1991kc}:
\be\label{eq:fazekasAF}
E_{AF}(x)=-\frac{1}{\pi}\int_{0}^{\pi} \sqrt{\frac{x^2}{16}+4\sin^2(k)}dk.
\ee

It is evident that the Hamiltonian (\ref{eq:MFHamiltonian}) contains all this physics and therefore its study (without forcing perfect antiferromagnetism)
will improve this RKKY-focused description. In particular it will permit to study how the Kondo effect and the RKKY-effect relate to each other. However
it is our opinion that other paths, rather than the mean field analysis, could also be followed; see Ref. \onlinecite{ChristmaMajorana,JohanKondoORIGINAL}
for examples of such a study.
Indeed at half-filling the most important feature that should be captured is the global singlet nature of the ground state; feature that would
be completely lost in any mean-field analysis that breaks translational invariance. Moreover away from half filling, the incommensurability of the order would
imply an increase of the unit cell used, bringing quickly to an untreatable form for the mean-field problem.

For all these reasons, and because the focus of the present work is on the ferromagnetism in the one-dimensional KLM, we will
not analyze these cases, but we will use the known results of the Ne\'el ordered ground state to make a comparison and complete our analysis.
As can be seen in Fig.~\ref{FIG:fazekasVSus}, the KI state has a better energy, respect to the antiferromagnetic ordered state, already at $x\approx 1.9$.

\section{Conclusions and outlook}
In this paper we have introduced a alternative representation of the Kondo Lattice model, in terms of three
spinless fermions interacting on a lattice. The identification of this map demonstrates by direct inspection the
known\cite{JohanKondoORIGINAL} representation of the KLM in terms of six Majorana fermions; moreover it
generates a Hamiltonian that is very suitable for the analysis of ferromagnetism in the one-dimensional Kondo lattice.

We performed such an analysis and showed how, already at mean-field level, many properties of the phase diagram
could be detected and explained. This is made possible by the identification of a symmetry of the Hamiltonian
that is responsible for the appearance of the ``commensurability parameter''.

Our work considerably improves the available mean-field analyses and is consistent with some recent
results obtained\cite{Peters:2012fc} by DMRG calculations, on the nature of the ferromagnetic metallic phase at intermediate and low couplings. In particular
we find the same value for the commensurability parameter and identify the same description of the system in terms of ``minority'' and ``majority'' electrons,
together with the emergence of the spin-selective Kondo insulator (SSKI), reported in recent studies\cite{Peters:2012fc}.

We showed how the existence of the SSKI stabilizes the ferromagnetic FM-I phase in the low density sector of the phase diagram.
At couplings $x\lesssim 2$ the system is in the FM-I phase only for densities $n_c<n^{F}_{crit}(x)$, while is is in the FM-II phase for
$n^{F}_{crit}(x)<n_c<1$.

The phase transition line $n^{F}_{crit}(x)$ lies reasonably close to the ferromagnetic-paramagnetic transition line,
signaling correctly the instability of the SSKI mechanism for excessively high densities.
The FM-II phase, that takes over beyond this transition, was instead identified as solution related to the RKKY-ferromagnetic one,
but optimized in order to capture more of the Kondo physics. This phase is present in the region of the phase diagram typically
occupied by paramagnetic ground states, for $x\lesssim2$. At low coupling it is energetically
more competitive than the
the usual mean field states with spiral spin order. This means that our result could, in principle, be further improved considering
modulations of the mean fields.
Existing up to half filling and for any coupling $x\lesssim2$, the FM-II phase represents a valid prototype for the ferromagnetic tongue
phase\cite{McCulloch:2002ss} that exists inside the ferromagnetic dome.
To the best of our knowledge there exist no other (not fully numerical) approach able to justify the existence of a ferromagnetic phase
in correspondence of the ferromagnetic tongue.

At coupling larger than $x\approx2$ the FM-II phase disappears and a region of phase coexistence between the FM-I phase
and the half-filled Kondo insulating one appears. We believe that such a region is due to the failure of the hypothesis done in the mean field decomposition
of the Hamiltonian. However the physical picture described by our results is not in contrast with the known physics of the KLM.
Moreover it suggests a qualitative change in the properties of the ferromagnet for high couplings.

At half-filling we discovered another translational invariant solution (that exists as a singular point in the phase diagram). We identify
this solution with the Kondo insulating one, recognizing that asymptotically it converges to the correct ground-state, with the correct
coupling dependence for the energy of the ground-state and for the quasiparticle gap. However, for small couplings,
it is not a good trial ground-state. We do not accomplish in this manuscript any
detailed analysis of the half-filled solutions, that instead have been the subject of a different study. In the
present work we simply identify the relation between the usual spiral ordered approximate solutions and our own.

Considered the great amount of physics, and the qualitatively convenient pictures that we have been able to elaborate, we hope
to have demonstrated to the reader the convenience of the analysis of the KLM in terms of Majorana fermions. In this work we used the
Majoranas to identify the three spinless fermions $c$, $g$ and $f$. We believe that the Majorana map is, in general, very advantageous
for the definition of these kind of non-linear fermion-spin mappings. A generalization of our approach, if appropriately used, can open the doors
towards a convenient description of a huge amount of unknown phenomena.

As final remark we would like to remind the reader that the $cgf$-map holds in any number of dimension. Therefore, differently
from bosonization or DMRG that find little use away from one dimension, our analysis can be straightforwardly applied
also in two and three dimensions.

\subsection{Acknowledgments}
\label{sec:conclusions}


We wish to thank the Swedish research council (Vetenskapsr{\aa}det) for funding.

\appendix

\section{Fermion representation}\label{subsec:kumar}

It is well known that in many condensed matter systems the electron does not behave as an elementary degree of freedom.
In recent years it became evident, both from theoretical
and experimental points of view, that under specific circumstances the collective electron modes, describing a normal Fermi liquid, can
decompose into
more fundamental
excitations with different quantum numbers and statistics\cite{Kim:2006ijBIS,*Chen:1992kx,*Schlappa:2011bsBIS,*Pen:1997oqBIS,*Shimojima:2011hcBIS},
e.g.\footnote{To this list we could include also the theoretically predicted pure Majorana excitations,
but being still absent an experimental decisive result, we omit them.} spinons, holons and orbitons.
The development of a formalism that does not focus on the quantum numbers of the electron and puts
aside its elementary nature could therefore be conceptually and formally advantageous. It is our opinion
that the best candidate for such a more elementary formalism is given by the Majorana representation of the quantum degrees of freedom.

Using {\it non-linear transformations}\cite{StellanMele}
on the local Fock space, it becomes straightforward to represent the fermion creation/annihilation
operators of a spinful electron in terms of {\it composite holon-spin operators} \cite{MatsStellanHUBBARD}.
The analysis of this kind of transformations turns out to be quite natural in terms of
Majorana fermions\cite{KumarORIGINAL}. A comprehensive discussion on these aspects will be the subject of another study\cite{MBzNilssonPREP},
so in this appendix we outline some known practical results that have been already used in the analysis of
the Hubbard model\cite{MatsStellanHUBBARD,KumarORIGINAL, StellanMele} and that will be useful in our main discussion.

In the literature on Majorana fermions the most used representation of spinful electron operators is
given by Eq.~(\ref{eq:cupmajorana})-(\ref{eq:cdownmajorana}). Another equivalent one\cite{KumarORIGINAL} is
\be
c^\dd_\ua&=&-\sqrt{2}\Phi \sigma^+ ,\label{eq:cupmajoranaKUMAR}\\
c^\dd_\da&=&\frac{2\Phi\sigma_z+i\Psi}{\sqrt{2}},\label{eq:cdownmajoranaKUMAR}
\ee
where $\Phi$, $\Psi$ are Majoranas and $\sigma^+=(\sigma_x+i\sigma_y)$, $\sigma^-=(\sigma_x-i\sigma_y)$ and $\sigma_z$ are the
usual Pauli operators, with the convention $\sigma^2_z=1/4$.
The relation between (\ref{eq:cupmajorana})-(\ref{eq:cdownmajorana}) and (\ref{eq:cupmajoranaKUMAR})-(\ref{eq:cdownmajoranaKUMAR}) is
given by the identifications:
\be
\Phi=2i\gamma_1\gamma_2\gamma_3,&&\qquad \Psi=\gamma_4,\\
\sigma^x=-i\gamma_2\gamma_3,\quad \sigma^y&=&-i\gamma_3\gamma_1,\quad \sigma^z=-i\gamma_1\gamma_2.\label{eq:kumarspins}
\ee
This representation for the creation/annihilation operators of the spinful electron
realizes the decomposition of the electron into its spinonic and holonic
components\cite{MatsStellanHUBBARD,KumarORIGINAL}, given respectively by the three Pauli operators
$\sigma_i$ and by the spinless fermion with creation operator $h^\dd=(\Phi+i\Psi)/\sqrt{2}$.
With these definitions it is possible to see that there exists a one-to-one correspondence between the
Hilbert space generated by the operators (\ref{eq:cupmajorana})-(\ref{eq:cdownmajorana})
starting from the vacuum state $|0\rangle$ such that $c_\da|0\rangle=c_\ua|0\rangle=0$, and the Hilbert space generated by the operators
$\{1,h\}\otimes \{\sigma^+, \sigma^-\}$ and their hermitian conjugates, where $h |0_h\rangle =0$
and $\sigma^+|\Uparrow\rangle=\sigma^-|\Downarrow\rangle=0$.
The mapping is given schematically in Tab.~\ref{tab:Kumarmap}.

\begin{table}
\caption{\label{tab:Kumarmap}Mapping, as introduced in Ref. \onlinecite{KumarORIGINAL}, between the two different representations of the Hilbert
space associated with a local spinful electron. On the left
the spinor representation, given by the operators  $c_\da$, $c_\ua$ and hermitian conjugates;
on the right the representation given in terms of holon and Pauli operators.}
\begin{ruledtabular}
\begin{tabular}{rcr}
\qquad$|0\rangle$&  $\longleftrightarrow$ & $|0_h\rangle\otimes|\Downarrow\rangle$\qquad \,\\
\qquad $|\ua\da\rangle$& $\longleftrightarrow$ & $|0_h\rangle\otimes|\Uparrow\rangle$\qquad \,\\
\\
\qquad $|\ua\rangle$&  $\longleftrightarrow$ & $|1_h\rangle\otimes|\Uparrow\rangle$\qquad \,\\
\qquad $|\da\rangle$& $\longleftrightarrow$ & $|1_h\rangle\otimes|\Downarrow\rangle$\qquad \,\\
\end{tabular}
\end{ruledtabular}
\end{table}

Two comments are in order on the operators $h^\dagger$ and $\sigma$. First of all it is remarkable that the Pauli operators (\ref{eq:kumarspins}) have to be
interpreted as spin or (charge) pseudospin operators,
depending upon the presence or the absence of the holon associated to $h^\dd$.
Secondly it is appropriate to note that the spinless-fermion creation operator $h^\dd$ is obtained via a transformation that
mixes the original Majorana $\gamma_4$ with the composite Majorana $2i\gamma_1\gamma_2\gamma_3$.
This gives an immediate understanding of the connection between the Hilbert space
of the 1-site Anderson model and the Hilbert space of the 1-site Kondo lattice model described in Sec. \ref{sec:majoranamap}:
in fact the confining term $U(n_f-1)^2$ is easily rewritten as $U(1-\phi^\dd \phi)$, where $\phi$ is the
holon associated to the spinful electron described by $f^\dd$ in (\ref{eq:andersonhamiltonian}).
Consequently also the origin of the operators (\ref{eq:kondospins}) becomes clear: for $U\rightarrow+\infty$ there must be one holon $\phi$ per site, that means
one spinful $f$-electron per site; therefore the operators (\ref{eq:kondospins}) must be interpreted as the spin operators of the original $f$-electron.
%
%
%
%


\section{Generalized algorithm for mean-field analysis of non-quadratic Hamiltonians}\label{app:A}

We outline the numerical procedure that we have used to study our system.
The method is not original\footnote{Stellan  {\"O}stlund, private notes (unpublished).}, but being unpublished it requires a quick (although not complete) introduction.

Mean-field theories are variational theories where the variational parameters are the mean-fields (order parameters) and the Hilbert space is
given by the states that can be written as a single Slater determinant of {\it properly defined} one-particle states.
The quantity that has to be minimized is the Free energy. In particular
a well known theorem\cite{Feynamn} says that
\be\label{eq:feynman}
F_{\text{true}}\leq F_{\text{trial}},
\ee
where $F_{\text{true}}$ is the Free energy associated to the density matrix of the original Hamiltonian $H$. A mean-field solution extremizes the 
unction $F_{\text{trial}}$, with respect to small variations of the mean field parameters.

Since the mean-field result is expressible as a Slater determinant, it means that there must exist a quadratic hamiltonian $\tilde H_{\text{mf}}$ generating the mean-field one-particle states;
consequently there must exist a mean-field density matrix $\tilde \rho_{\text{mf}}=\exp(-\beta \tilde H_{\text{mf}})/\tilde Z$, so that
\be
F_{\text{trial}}=\text{Tr}\left(\tilde \rho_{\text{mf}}H\right)-TS_{\tilde\rho_{\text{mf}}}.
\ee
In second quantization terms this Hamiltonian must take the form
\be
\tilde H_{\text{mf}} = \sum_i \mu_i A_i,
\ee
where the $\mu_i$ are (real) parameters that we call {\it variational parameters} and the $A_i$ are {\it all} the possible quadratic (Hermitian) operators, written in terms of the original particle creation/annihilation
operators that appear in $H$. For future convenience the parameters $\alpha_i=\langle A_i\rangle=\text{Tr}(\tilde \rho_{\text{mf}} A_i)$ are named {\it order parameters}.
It is evident that
the specific value of any order parameter $\alpha_j$ will (in general) depend on the entire set $\left\{\mu_i\right\}$.

This means that the term $\text{Tr}\left(\tilde \rho_{\text{mf}}H\right)$ will correspond to the mean-field energy functional that one can obtain via Wick decomposition of all the operators that belong to $H$.
So
\be
\langle H\rangle=\text{Tr}(\tilde \rho_{\text{mf}} H)=\mathcal H(\alpha_i),
\ee
that implicitly means also $\mathcal H(\mu_i)$. Of course the same can be said for the term $TS_{\tilde\rho_{\text{mf}}}$ that becomes $T\mathcal S(\mu_i)$.

The best mean-field solution is given by the density matrix that minimizes (\ref{eq:feynman}), but in general all the solutions that extremize it are acceptable mean-field solutions.
Of course
the best one will be the one with lowest Free energy.
Extremizing $F_{\text{trial}}(\mu_i)=\mathcal H(\mu_i)-T\mathcal S(\mu_i)$ one obtains the condition:
\be
0=\left(\frac{\partial \mathcal H}{\partial \alpha_i}-\mu_i \right) \frac{\partial \alpha_i}{\partial \mu_j}.
\ee
Unless it happens that there exists an $\alpha_i$ independent of all the $\{\mu_j\}$, then one must have
\be\label{eq:mainselfcons}
\frac{\partial \mathcal H}{\partial \alpha_i}=\mu_i.
\ee
This is a set of (non-linear) equations in the parameters $\{\mu_i\}$; clearly there exists one equation per $\mu_i$.

The algorithm is then implemented in a straightforward way:
\begin{enumerate}
\item Given the original Hamiltonian $H$, one has to start writing down all
the possible (not necessarily hermitian) pairings of creation/annihilation operators that appear in $H$, generating a set of possible
order parameters $\beta_i$ (note that this set can be infinite in principle, because it can contain also very non-local order parameters).
An analysis of the set $\{\beta_i\}$ must be done, inserting the information about the physics: for example symmetries, continuous or discrete,
that have to be preserved (for example translational invariance, or time-reversal symmetry)
or conditions given by the hermitean character of the Hamiltonian (for example if $\langle c^\dd g^\dd\rangle=\Delta$ then
$\langle c g\rangle=-\Delta^*$ by hermiticity).
This will create the set of order parameters $\{\alpha_i\}$, in general smaller (always finite), introduced previously.

\item Given the different order parameters $\alpha_i$ one writes down the operators $A_i$ that correspond to them, such that $\langle A_i\rangle=\alpha_i$.
Note that these operators can be (properly normalized) linear combinations of quadratic operators.

\item Given $H$ and the set $\{\alpha_i\}$ one can write down the mean-field functional $\mathcal H(\alpha_i)$ obtained by standard Wick decomposition.

\item Using (\ref{eq:mainselfcons}) one writes down the (non-linear) system in terms of the variation parameters $\mu_i$.
This non linear system can then be solved numerically. And its solutions are by construction also mean-field solutions of the Hamiltonian $H$.
\end{enumerate}

Although there are some physical interesting features hidden in this method, we will not to comment on it further here.
The method, treating all the mean-fields on equal footing, proved itself quite good in the study of the competition between different order parameters.

\section*{References}
\bibliography{Biblio2}

\end{document}